# Driving a stimuli-responsive wedge in the packing of phospholipid membranes using bolaamphiphile intercalants

Niki Baccile,[a,*] Archan Vyas,[b] Ramanujam Ramanujam,[a] Daniel Hermida-Merino,[c] Ingo Hoffmann,[d] Lionel Porcar,[d] Atul N. Parikh[b,e,*]

[a] Sorbonne Université, Centre National de la Recherche Scientifique, Laboratoire de Chimie de la Matière Condensée de Paris, LCMCP, F-75005 Paris, France

[b] Departments of Biomedical Engineering and Chemical Engineering, University of California, Davis, California 95616, USA

[c] Departamento de Física Aplicada, CINBIO, Universidade de Vigo, Campus Lagoas-Marcosende, Vigo, 36310, Spain

[d] Institut Laue-Langevin, 38042 Grenoble, France

e School of Materials Science & Engineering, Singapore Centre for Environmental Life Sciences & Engineering, and Institute for the Digital Molecular Analytics & Science, Nanyang Technological University, Singapore

*** Corresponding author:**

Dr. Niki Baccile

niki.baccile@sorbonne-universite.fr

Prof. Atul Parikh

anparikh@ucdavis.edu

**Abstract**

Bolaamphiphiles – amphiphilic molecules with polar groups at each of the two ends of a hydrophobic tail with pH-sensitive spontaneous molecular curvatures – endow membranes of extremophiles with an exquisite balance between stability (or robustness) and adaptability (or plasticity). But how the presence (or real-time insertion) of bolaamphiphiles influences lamellar lipid membranes is poorly understood. Using a combination of time-resolved confocal fluorescence microscopy, *in situ* small angle X-ray and neutron scattering (SAXS, SANS), and neutron spin echo (NSE) measurements, we monitor here the pH-dependent interactions of nanoscopic vesicles of a representative bolaamphiphile – a glucolipid consisting of a single glucose headgroup and a C18:1 (oleyl) fatty acid tail (G-C18:1) – with the membranes of an essentially cylindrical, fluid-phase phospholipid (dioleoylphosphatidylcholine, DOPC). We find that the two mesophases interact spontaneously at all pH values, producing large-scale morphological remodeling. Under neutral and acidic conditions, when the bolaamphiphile assumes a cylindrical shape, vesicles fuse with one another, producing invaginations, inner tubulation and vesicle-in-vesicle aggregates. Under basic pH, by contrast, when the carboxylic acid is deprotonated and the molecule is inverted-conical in shape, the bolaamphiphile causes phospholipid membranes to undergo poration, budding, and vesiculation. This pH-dependent, environmentally sensitive membrane remodeling without the disruption of the essential bilayer motif illustrates how local, molecular-level packing perturbations can translate into global system-level morphological changes, enabling membranes to acquire environmental sensitivity and real-time adaptability. These results support the notion that molecular fluxes – which add (or remove) amphiphilic molecules to biological membranes – can endow *de novo* functionalities (e.g., pH sensitivity) and influence global morphologies of cell-sized vesicles.

**Introduction.**

Many common phospholipids – the primary molecular constituents of the bilayer component of the cellular membranes – are essentially cylindrical.[1,2] At mechanical equilibrium, they assemble into lamellar bilayers with negligible spontaneous curvature $(C_o)$, which is the equilibrium curvature their monolayers adopt in the absence of constraints.[3,4] The insertion of non-cylindrical amphiphiles with non-zero spontaneous curvatures drives a wedge in the membrane, locally disrupting the intermolecular packing.[4,5] For example, the incorporation of *inverted-cone*-shaped amphiphiles (lysolipids, detergents, and surfactants), which have bulky hydrophilic heads compared to their hydrophobic tails confer positive curvatures $(C_o > 0)$, causing the monolayer to bulge towards the headgroups. As a result, the accumulation of inverted-cone-shaped amphiphiles drives a shape transition from the lamellar (*lam*) to tubular ($H_I$) or spherical micelles. Conversely, *cone-shaped* molecules like diacylglycerols, oleic acid, and phosphatidylethanolamines, with smaller heads and wider tails, locally promote the generation of negative curvatures $(C_o < 0)$. This, in turn, favors the morphological shift from lamellar to the inverse hexagonal ($H_{II}$) or cubic (bicontinuous) phase.[6] The generation of these spatially localized curvatures and accompanying shape changes are, however, opposed by the global lamellar motif, which produces elastic curvature stress – a free energy penalty – in the lamellar membrane.[6,7] Beyond a critical concentration of non-cylindrical wedge molecules, these elastic stresses become manifest, inducing local and global shape deformations (e.g., budding, invagination, and tubulation) and, in certain cases, topological transformations (i.e., fusion and division).[6] In the limiting cases, the lamellar motif is abandoned and replaced by solubilized fragments.

An interesting feature of some of these wedge amphiphiles is that their molecular shapes are environmentally tunable. They thus perturb the lamellar membranes in characteristic manners dependent on the state of the environment. This is perhaps best exemplified by the pH-dependent incorporation of single-chained fatty acids (FAs) into lamellar membranes of double-chained phospholipids. At low pH, the cone-shaped geometry of the small, neutral headgroups (COOH) and voluminous hydrocarbon tails promote negative spontaneous curvature (e.g., inverted hexagonal $H_{II}$ phase)[8–10] leading to *Lam*-to-$H_{II}$ transitions[11]. By contrast, at elevated pH, the deprotonation of the headgroups ($COO^-$) changes the molecular shape from conical to cylindrical or inverted-conical shape, thus favoring positive curvatures and leading to micellar or lamellar structures.[12] These pH-dependent shape transitions of FA-containing phospholipid membranes can thus serve as discriminating reporters of the environmental acidity. When the environmental pH is lowered below neutrality, the membrane curvature becomes progressively negative (cone-like effect) until phase separation ($H_{II}$) occurs below pH 6.[13–15] Above neutral pH, FAs adopt surfactant-like behavior (inverted-cone effect), which disrupts the membrane producing worm-like mixed micelles.[15]

The considerations above raise several general questions: how would the bolaform molecules – those amphiphiles that feature polar groups at both ends of a hydrophobic tail and organize into monomolecular layers (rather than bilayers)[16] – interact with the bilayer-forming phospholipid membranes? How do these interactions help stabilizing membranes of extremophile organisms under conditions of extreme environmental stresses (e.g., pressure, temperature, or desiccation)?[17–22] How does the local environmental sensitivity (i.e., pH dependence) of bolaamphiphiles influence the global morphology of membranes doped with them? Addressing these questions will yield important insights into understanding their contributions to the structural stability and environmental adaptability of membranes of living organisms, like extremophiles.

To begin addressing these questions, we consider here membrane interactions of the class of natural fatty acid-based bolaamphiphiles known as biosurfactants[23]. These biologically-derived amphiphiles are characterized by one or more FA covalently linked to a carbohydrate (e.g., sophorose, rhamnose, mannose, cellobiose, xylose), generally at the ω or ω-1 end-group. Much of the existing effort in characterizing the interactions between biosurfactants and phospholipid membranes[24] is limited to rhamnolipids (RLs), a related class of gemini[25] amphiphile. The cumulated weight of these studies establish that the RLs interact with phospholipid membranes through the wedge effect. Specifically, as inverted-cone amphiphiles, they counterbalance the conical shape of dielaidoylphosphoethanolamine (DEPE) lipids and stabilize a flat membrane with negligible spontaneous curvature or $C_o$.[26] In this same vein, their incorporation into flat membranes of essentially cylindrical phosphatidylcholine (PC) lipids[27,28] stabilize a net spontaneous curvature producing buds, protrusions, filaments, or even induce topological transitions such as fission.[29–32] Like many FAs, RLs also exhibit a rich pH-dependent phase behavior[25], but the influence of pH on curvature synergy between RLs and PC membranes has not been studied. Moreover, interactions of biological membranes with other biosurfactants, such as sophorolipids, cellobioselipids or glucolipids, which exhibit rich pH dependence and environmental sensitivities have yet to be fully studied.[24]

Here, we study the interactions between a structurally simple (linear) bolaamphiphile, a single-glucose derivative of oleic acid (G-C18:1),[33] and a prototypical phospholipid membrane consisting of a fluid phase, namely dioleoylphosphatidylcholine (DOPC, **Figure 1a**). G-C18:1 is known to exhibit a rich pH dependent self-assembling properties (**Figure 1b).**[34,35] The molecule combines a FA- and glucolipid-like behavior, it assembles into monolayers characterized by $C_o \sim 0$ under acidic (pH < 6.5) and micelles ($C_o > 0$) under basic (pH > 7) conditions.[35] We monitor in real-time the effect of the presence of G-C18:1 vesicles on cell-sized giant unilamellar vesicles (10-25 micrometers in diameter) composed of DOPC. Evidence from optical fluorescence microscopy and X-ray scattering measurements reveal that G-C18:1 interacts readily within the DOPC mesophase, by the mechanism of vesicle fusion. The interaction induces large-scale morphological remodeling inducing mesoscopic budding, tubulation, and intra-vesicle vesiculation. At the molecular level, X-ray evidence supports uniform thinning. Moreover, the incorporation of G-C18:1 into DOPC bilayers introduces pH sensitivity: the pH dependence of molecular shape of G-C18:1 is translated at the larger morphological length scale. At elevated pH – when deprotonated G-C18:1 adopts an inverted-cone shape – interaction with the phospholipid bilayer stabilizes positive curvature deformations (budding, outer vesiculation). By contrast, at lower pH, when G-C18:1 is cylindrical in shape, the association promotes aggregation of phospholipid vesicles. These results support the notion that molecular fluxes – which add (or remove) amphiphilic wedge molecules to biological membranes – can endow *de novo* functionalities (e.g., pH sensitivity) and influence global morphologies of cell-sized membranous structures.

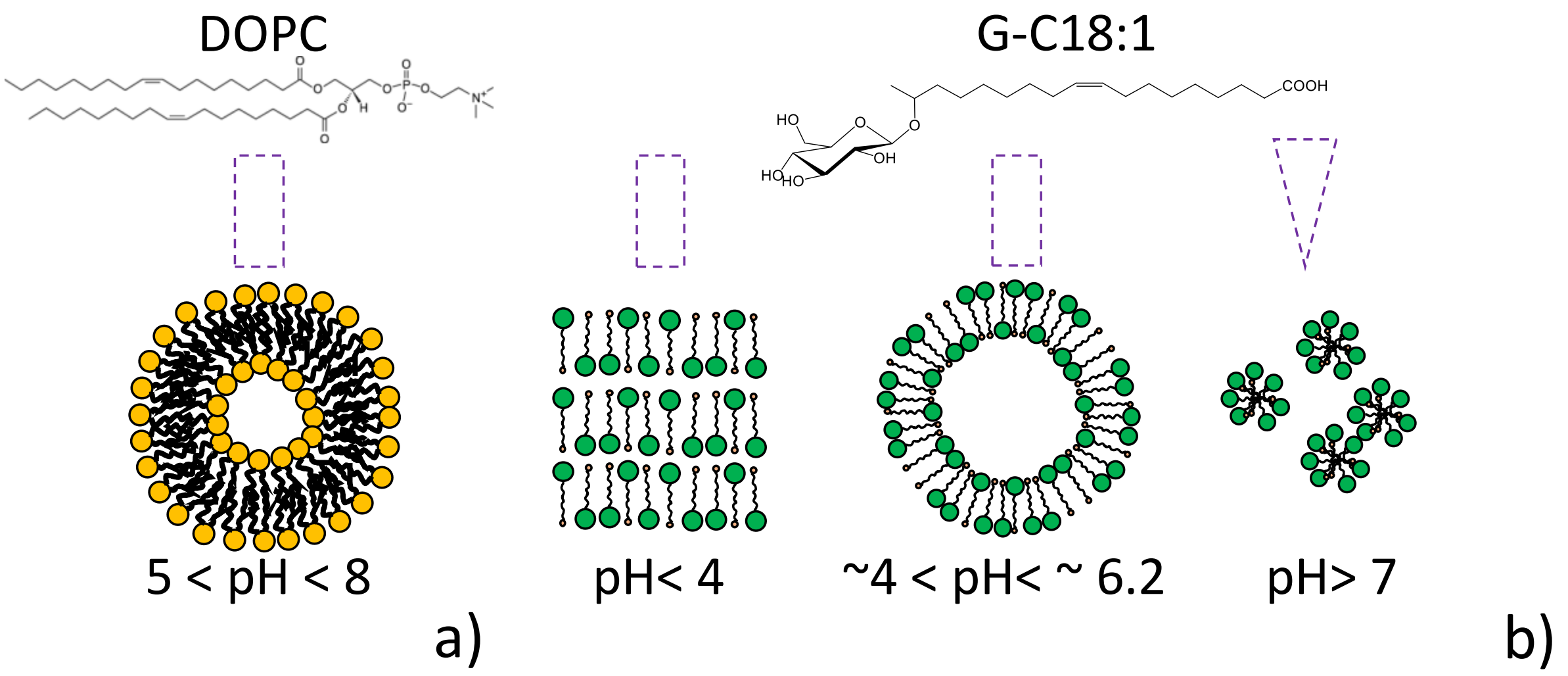

**Figure 1 – Lipids used in this work. Their known self-assembled structures, and the corresponding molecular shape (cylinder, inverted cone), in water are reported for various pH ranges.**

## Results and discussion

We begin by preparing vesicular mesophases of the two amphiphiles under a range of pH conditions (See **Experimental Methods** section). DOPC GUVs are prepared by following the standard electroformation technique stained in red by a common rhodamine-modified DOPC lipid dye (**Figure S 1**). Over the entire pH range examined (3 < pH < 10), the well-formed GUVs display an essentially spherical and weakly undulating membranous boundary (**Figure S 2, Figure S 3**), in agreement with previous studies.[36] Note that, in addition to GUVs, electroformation inevitably also produces a diversity of mesophases including multilamellar and nested giant vesicles. For our study, we focus our observations on GUVs with empty lumen alone. By contrast, our efforts to prepare GUVs using G-C18:1 (**Figure S 2**) yielded fragile vesicles, which frequently unfolded during handling, ultimately producing ill-characterized lipid suspension. For this reason, we adopt a different hydration method (dispersion in water, pH adjustment, and bath sonication), which produces an order-of-magnitude smaller vesicles ($100\ nm < d < 1$ µm, **Figure S 2**) between pH 4 and 6.2.[37,38] The G-C18:1 vesicles are stained with a strongly hydrophobic blue dye (perylene, **Figure S 1).**

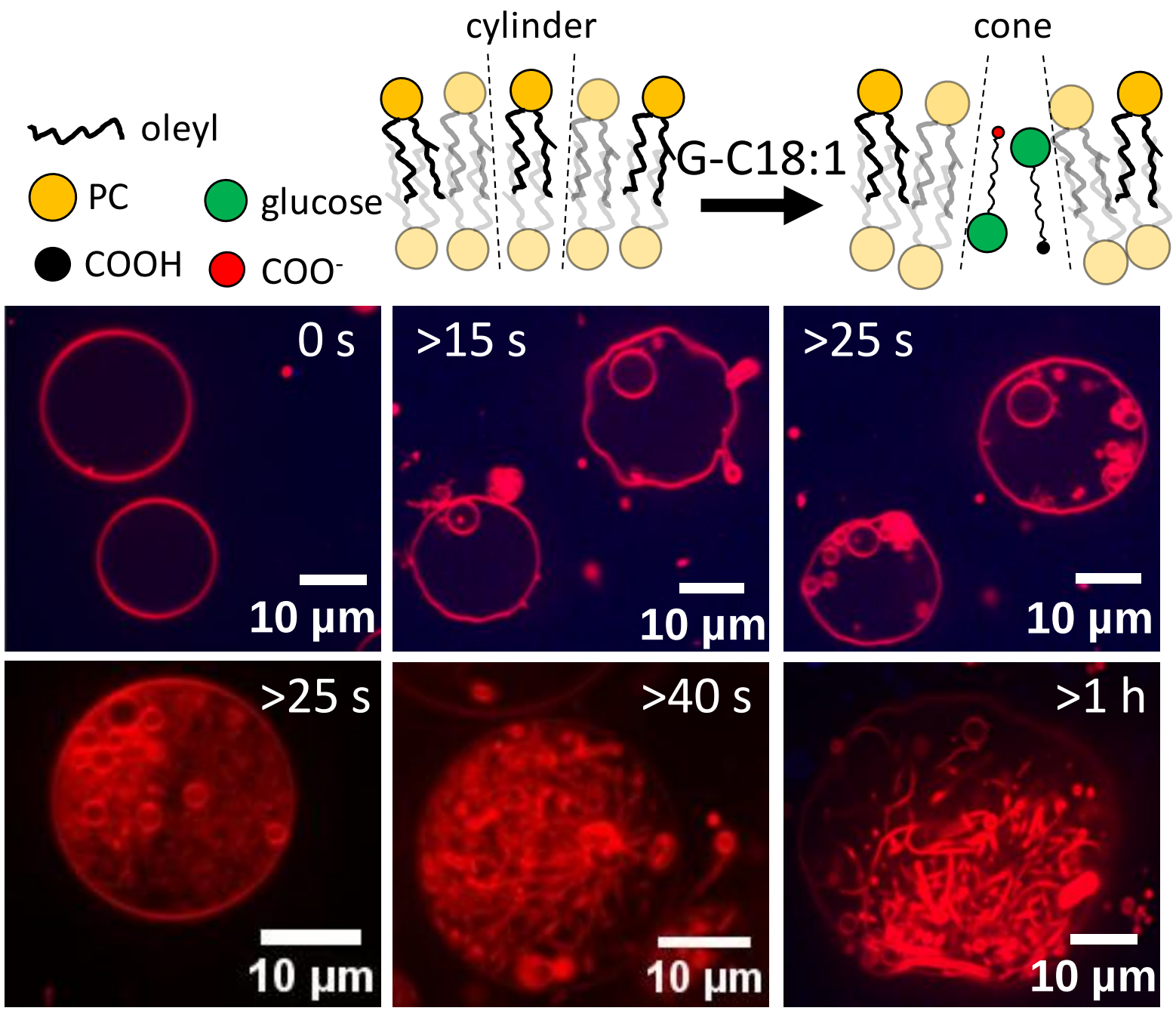

**Figure 2 – Representatives screenshots showing real-time deformation of DOPC GUVs (0 s) after interacting with a suspension of glucolipid G-C18:1 vesicles. Temporal images at >15 s and >25 s show vesicles in vesicles motifs and are extracted from Video S 2 for $C_{G\text{-}C18:1}$= 0.5 mg/mL, pH 5.5. Images >40 s and >1 h focus on inner tabulation and are extracted from Video S 0 and beginning of Video S 3, respectively, with $C_{G\text{-}C18:1}$= 5 mg/mL, pH 5.5. Extended experimental conditions: Table 1 in the Experimental Methods section.**

**The cone-like arrangement of G-C18:1 in the cartoon is deduced from the invagination events observed after 15 s.**

Mixing pre-formed, osmotically balanced, G-C18:1 single unilamellar vesicles (SUVs) with DOPC GUVs prompts large-scale shape changes even in nominally spherical GUVs. Within tens of seconds and after 1 h after mixing the two mesostructures, a diversity of shape deformations becomes visible (**Figure 2, Supplementary Video S 0 to Video S 3**). Some of the most frequent shape changes include the appearance of microscopic buds, pits and, above all, tubular invaginations. In addition, we often observe unusual "vesosome" structures consisting of nested arrays of smaller microscopic vesicles. Qualitatively, these shape transformations were fully reproducible in multiple independent experiments (n= 3) and at two different concentrations of G-C18:1 (**Figure 2, Video S0 to Video S3**). Additional complementary experiments (**Figure S 4** to **Figure S 8**) further confirm the interaction and fusion between DOPC GUVs and G-C18:1 vesicles.

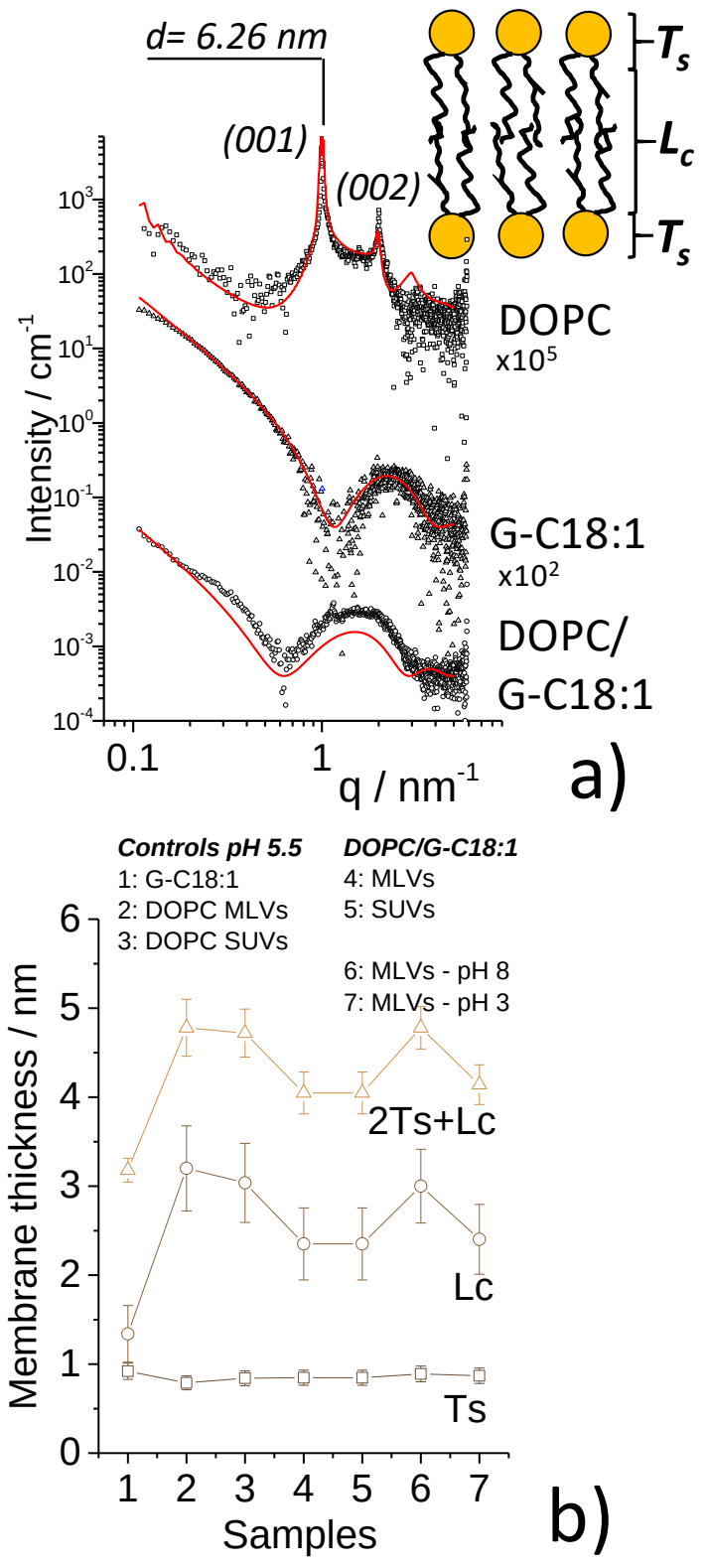

**Figure 3 – Probing the membrane structure by SAXS. a) SAXS profiles recorded on a G-C18:1 vesicle suspension (5 mg/mL, pH 5.5), DOPC MLVs (2 mg/mL, pH 5.5) and DOPC/G-C18:1 of molar ratio [DOPC]:[G-C18:1]= 0.8 ($C_{DOPC}$= 2 mg/mL, $C_{G\text{-}C18:1}$= 1.4 mg/mL, pH 5.5) after 15 min. Details about modeling SAXS data are provided in the Supporting Information. b) Bilayer membrane thickness estimated from modelling SAXS data.**

To characterize the interactions between the wedge molecules and the membrane at the finer mesoscopic length scales (< 50 nm), we carried out static and time-resolved *in situ* synchrotron SAXS measurements. Aqueous suspensions of GUVs in our hand did not produce sufficient SAXS signal to enable unequivocal characterization. To circumvent this issue, we replaced GUVs with multilamellar vesicles (MLVs) and small unilamellar vesicles (SUVs) prepared by the hydration method (refer to the **Experimental Methods section**). Compared to SUVs, we note that MLVs have additional structural determinants, such as interlamellar interactions and suppressed fluctuations. Thus, the correspondence of results obtained between the two systems requires careful consideration. However, in the main text, we highlight the results obtained on MLVs, because focusing on the impact of G-C18:1 vesicles on the (001) lamellar peak avoids the use of numerical models both in static (**Figure 3**) and *in situ* (**Figure 4**) experiments. The data recorded on SUVs are nonetheless presented on **Figure S 9**. Also worth to be noted that some numerical models may not perfectly match the experimental data, specifically after mixing DOPC and G-C18:1. This is inevitable and explained by the coexistence of structures, which would require multiple models and a large set of free, unconstrained, parameters. To avoid so, it was priviledged to employ one single (lamellar) model and match at best the oscillation of the form factor, of which the shift in the minimum can reasonably be attributed to a change in the membrane thickness.

The SAXS data of unperturbed DOPC multilamellar vesicles (MLVs) display the prominent (001) and (002) lamellar peaks (DOPC, **Figure 3a**), which correspond to the d-spacing of 6.26 nm, in good agreement with previous reports of other PC systems.[27,28] When G-C18:1 vesicles (control SAXS given in **Figure 3a**) are added to the MLVs, the resulting minimum of the form factor shifts to higher q-values, suggesting thinner membranes (DOPC/G-C18:1, **Figure 3a**). Note also that the signature of the mixed sample is different from the SAXS profile of G-C18:1 alone (**Figure 3a**). Modelling all SAXS profiles (**Figure 3a**) with a standard head-tail lamellar model allows us to estimate the total membrane thickness, $T = 2T_s + L_c$, shown in **Figure 3b**, with $T_s$ the shell (head) thickness and $L_c$ the core (tail) length. These analyses indicate that roughly 15 min after adding the G-C18:1 vesicles, T measures 4.0 ± 0.4 nm. This reflects a -17% reduction from the thickness obtained for pure DOPC MLVs (T= 4.8 ± 0.5 nm) and a +25% increase in comparison with the thickness of G-C18:1 (T= 3.2 ± 0.3 nm) vesicles.[28,37,38] Discrimination between the head and tail components nicely shows that the change in thickness essentially depends on the tail length only (**Figure 3b**), a fact that can be qualitatively verified by the shift in the minimum of the form factor (**Figure 3a**). Similar results are obtained from the G-C18:1 in interaction with SUVs (3 and 5 in **Figure 3b**): T is reduced from 4.7 nm for DOPC SUVs to 4.1 nm when G-C18:1 is added.

Based on the results above, we can deduce that G-C18:1 influences DOPC MLVs in two major respects: it transforms the multilamellar DOPC vesicles into a unilamellar structure and thins the DOPC membrane. Note that the influence of oleic acid, a common single-chain amphiphile, on the d-spacings of structurally-related DEPE[39] and materially identical DOPC[40] membranes is practically negligible regardless of the pH. Even the effect of sophorolipids (0.1 mol %), a known bolaform biosurfactant similar to G-C18:1 but with two glucose units, on the d-spacing and membrane thickness of DPPC (6.09 nm) in its fluid phase (50°C) is also negligible (+0.02 nm only).[41] By contrast, the observed MLV-to-SUV transition and modulation in the PC membrane thickness is comparable to the previous reports of the effect of di-rhamnolipids (di-RLs) on DPPC membranes,[27,28] where the membrane thickness progressively reduces from 4.8 nm for pure DPPC to 2.5 nm for pure RLs.[28] We stress here that the MLV-to-SUV transition observed by SAXS after adding G-C18:1 to MLVs may not have direct parallel with GUVs, which have only one bilayer. However, the similar results found when G-C18:1 vesicles are added to DOPC SUVs (3 and 5 in **Figure 3b**) encourage the present approach and stimulated further experimental effort in setting a time-resolved *in situ* SAXS study on both MLVs (**Figure 4**) and SUVs (**Figure S 9**).

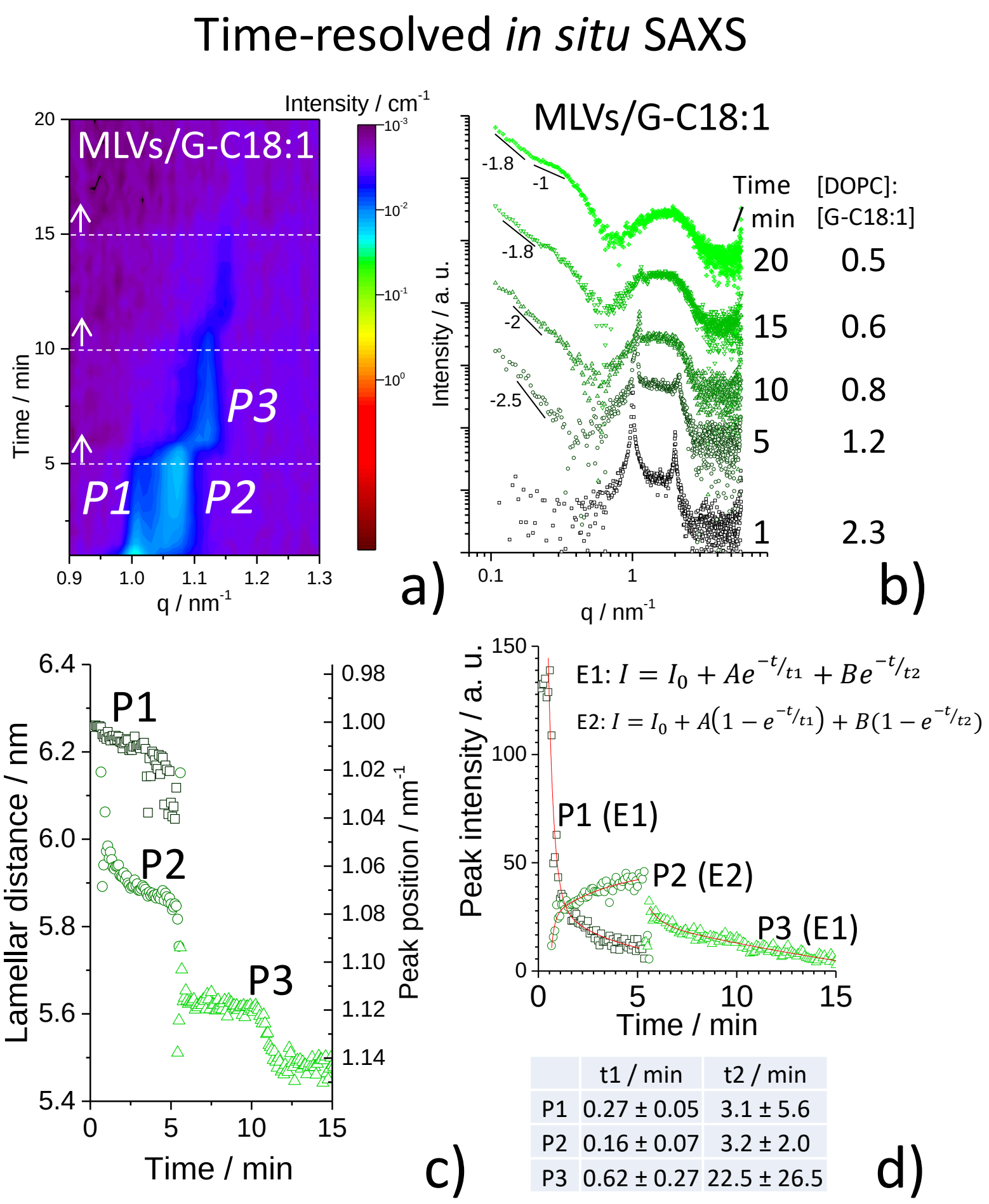

| | t1 / min | t2 / min |
|---|---|---|
| P1 | 0.27 ± 0.05 | 3.1 ± 5.6 |
| P2 | 0.16 ± 0.07 | 3.2 ± 2.0 |
| P3 | 0.62 ± 0.27 | 22.5 ± 26.5 |

**Figure 4 – Following membrane restructuring in real time. a) Time-resolved *in situ* SAXS: at t= 0, aliquots of the G-C18:1 vesicle suspension (20 mg/mL, pH 5.5) are added every 5 min to the DOPC MLVs (2 mg/mL, pH 5.5) suspension. The molar ratio [DOPC]:[G-C18:1] evolves from 2.3 at t= 0 to 0.5 at t= 20 min, while pH is 5.5. All quantitative details are given in Table S 1 in the Supporting Information. The (001) lamellar reflections are noted in P1, P2, and P3, and they are fitted over time using a Lorentzian function (position, intensity, full width at half maximum), of which the baseline is set at the bottom of the Bragg peak (contribution of the form factor oscillation is not taken into account). b) I(q)-representation of spectra at selected points in time and given molar ratio. c) Time resolved evolution of the peak position, $q_0$, and lamellar distance ($2\pi/q_0$) corresponding to P1, P2 and P3. d) Time evolution of the P1, P2, P3 peak intensities and fitted using a double exponential decay (E1, $I = I_0 + Ae^{-t/t1} + Be^{-t/t2}$) and growth (E2, $I = I_0 + A(1 - e^{-t/t1}) + B(1 - e^{-t/t2})$) functions.**

Time-resolved *in situ* SAXS data allow to follow and quantify the kinetics of the interactions between G-C18:1 and DOPC vesicles. The time evolution of the (001) lamellar peak and the low-q slopes (log-log scale, q< 0.8 nm$^{-1}$) of SAXS upon adding G-C18:1 (t= 0 s, [DOPC]:[G-C18:1]= 2.3) are presented both as a contour plot (**Figure 4a**) and a stack (**Figure 4b)**, while **Figure 4c** and **Figure 4d** present the time evolution of the d-spacing and peak intensity of the integrated (001) reflections. Three peaks (P) were identified. P1, initially the only (001) reflection and associated to DOPC MLVs, decays within the first five minutes after adding G-C18:1. P1 coexists with P2 (interlamellar distance smaller than P1 of about 0.3 nm), which grows concomitantly. Both P1 and P2 have comparable kinetics of exponential decay and growth, respectively, as evidenced by the comparable time constants, t1 (~0.2-0.3 min, that is 10-20 s) and t2 (~ 3 min), being comparable within the error but with t1 << t2 (**Figure 4d**).

After 5 min, an extra dose of G-C18:1 ([DOPC]:[G-C18:1]= 1.2) is added to the system. The close-to-stoichiometry composition stimulates a transition between P2 and P3, having a difference of about additional 0.3 nm, a similar two-step decay kinetics with t1< t2, although much slower (t1= 0.62 ± 0.27 min, about 30-40 s; t2= 22.5 ± 26.5 min). Further addition of G-C18:1 (t= 10 min, [DOPC]:[G-C18:1]= 0.8 and t= 15 min, [DOPC]:[G-C18:1]= 0.6) stabilizes the final inter-lamellar distance at 5.48 nm, but it does not influence the kinetics, before the MLV-to-SUVs transition. Overall, if the under-stoichiometric content of G-C18:1 has immediate effects on the DOPC membrane, most structural changes occur at and above the equimolar amount of G-C18:1.

The evolution of the slope (log-log scale) at q< 0.8 $nm^{-1}$ in **Figure 4b** is also notable, although to be taken with care, considering the narrow range of wavevector probed by the SAXS configuration. Typically, values contained between -1 and -2 suggest wormlike micelles (-1.6),[42,43] whereas -1 is generally observed for cylinders and -2 for bilayer membranes.[44] When G-C18:1 is added to DOPC MLVs, the slope evolves from -2 towards -1, with the coexistence of slopes after 20 min and total loss of the lamellar structure. This suggests the coexistence of two or more objects, possibly tubular and flat. The coexistence between families of objects is also supposed by the fact that modeling the DOPC/G-C18:1 SAXS profile after an equilibration of 15 min (**Figure 3a**) cannot be satisfactorily performed with one single form factor model (head-tail lamellar). Requirement of a second model is certainly necessary. However, in the absence of complementary microscopy data, and in light of the morphological complexity observed with confocal electron microscopy (**Figure 2**), use of simultaneous form factor models is not preferred. For comparison, neither FAs nor RLs (mono- or di-) are known to induce such drastic morphological transitions to PC membranes. According to SAXS arguments, RLs keep the DPPC membrane fluid by stimulating the formation of swollen liquid crystalline membranes.[28]

Next, we quantified the membrane dynamics, before and after addition of G-C18:1, using neutron spin echo (NSE) measurements associated with small angle neutron scattering (SANS), whereas SANS (**Figure S 10c**) is used to confirm the membrane morphology (low-q slope being -2) for both controls (G-C18:1, DOPC) and mixed DOPC/G-C18:1 samples. The DOPC control must necessarily be composed of SUVs, as otherwise the structure factor of MLVs interferes with the spin-echo scattering function (de Gennes narrowing).[45,46] NSE was chosen because it is the only method available to measure the scaled bending rigidty, $\frac{k}{k_b T}$ (Eq. 3 in Supporting Information), independently of the morphology and size of the scattering objects, while other common methods (flickering spectroscopy, pipette aspiration) require the use of GUVs.[47] By employing the classical pre-factor of 0.0069 (refer to discussion about NSE methods in the Supporting Information),[48,49] generally accepted for phospholipid GUVs, but probably questionable for more complex morphologies as in mixed DOPC/G-C18:1, it is found that of G-C18:1 vesicles moderately reduce (12%) the bending rigidity of DOPC vesicles (**Figure 5,** with the Zilman-Granek decay parameter $\frac{\Gamma_{ZG}}{q^3}$ and raw data given in **Figure S 10a,b**). Such moderate reduction is consistent with literature values, showing that low to moderate amounts of oleic acid lower the bending rigidity of cholesterol-free DOPC membranes by less than 10%.[40,50]

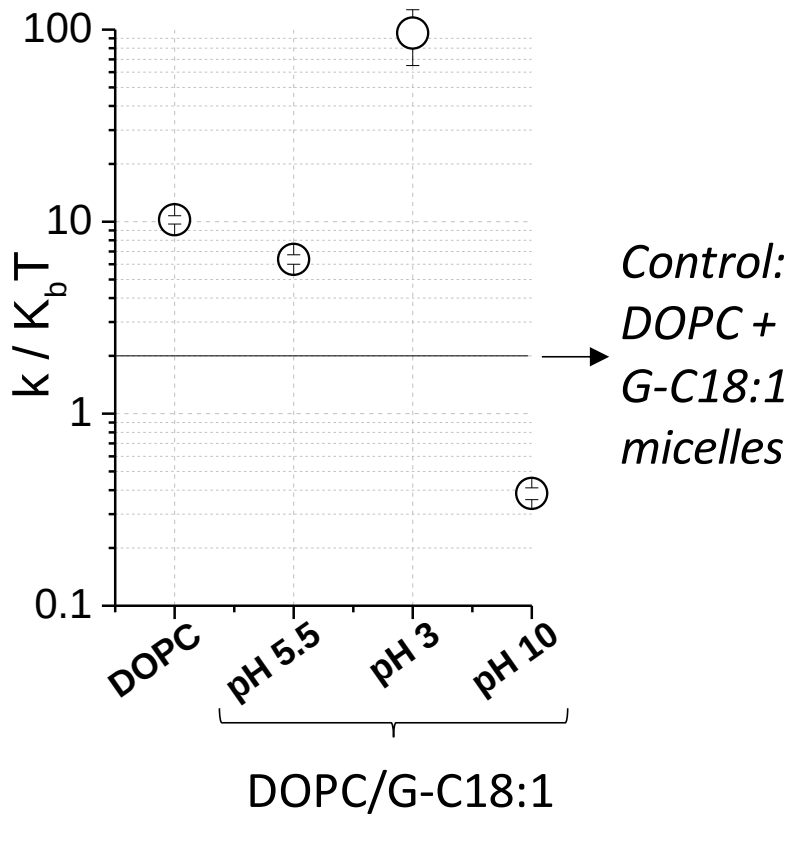

**Figure 5 – Probing membrane rigidity with NSE experiments in deuterated water. Scaled bending rigidity, $\frac{k}{k_b T}$, of DOPC SUVs, DOPC/G-C18:1 ([DOPC]:[G-C18:1]= 0.6) vesicles recorded at pD 6 and after jump to pD 3.6 and 10. The normalized, q-dependent, spin-echo intermediate scattering functions, Zilman-Granek decay parameter, and SANS experiments are provided in Figure S 10 in the Supporting Information.**

## pH dependent phase behaviors of bolaamphiphile containing phospholipid GUVs

pH triggers two different phase transitions in G-C18:1 vesicles (**Figure 1 b**): above pH 6.2, deprotonation shifts the shape of G-C18:1 from cylindrical to conical, with the result of inducing a vesicle-to-micelle transition.[35,37,38] pH below 4 induces the precipitation of a lamellar phase.[35,37,38,51] These transitions were reproduced in the DOPC/G-C18:1 fused vesicles with the aim of manipulating the morphology and structure of DOPC vesicles by means of G-C18:1.

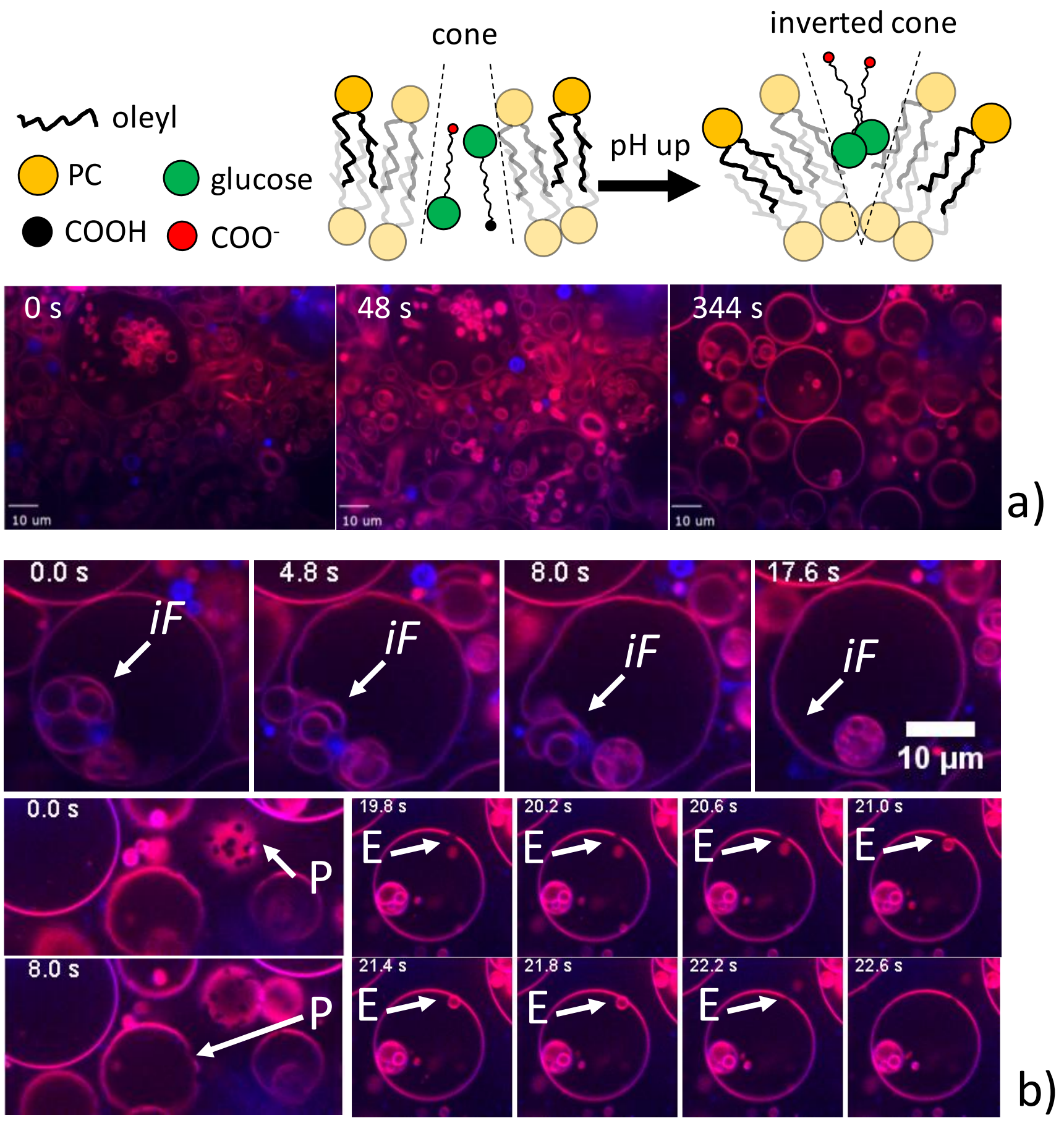

**Figure 6 – Effect of increasing pH on DOPC/G-C18:1 mixed systems presented in Figure 2. a) Time-resolved *in situ* evolution of an equilibrated (1 h) sample to which aliquots of a 0.1 M NaOH solution is added (experimental conditions: Table 1 in the Experimental Methods section). b) selected highlighted regions illustrating mechanisms of membrane modification: *iF*: internal fusion; *P*: poration; *E*: expulsion. The cone-to inverted cone-like arrangements of G-C18:1 in the cartoon are deduced from the exvagivation phenomena observed after increasing pH.**

Samples composed of DOPC and G-C18:1 are best described as GUVs containing inner vesicles and tubes, a structural heterogeneity and complexity accentuated by time: multiple features coexist, including nested and polydisperse vesicles of variable sizes, overall shapes, and complex structural hierarchies (**Figure 2a, Video S 0 to Video S 3**). The effect of adding a base (NaOH) to freshly prepared samples generates defects at the membrane surface (**Figure S 11, Video S 4**) resulting in the expulsion of inner content of the GUV and stabilization of well-defined GUVs. Similar changes are exacerbated in equilibrated (1 h) samples (**Figure 6, Figure S 11, Figure S 12)**: the initial morphological complexity disappears in time until GUVs are again the main stable objects (**Figure 6a)**. The close-up images in **Figure 6b** recorded at different locations in the 1 h equilibrated sample elucidate the mechanism and give the time scale of the phenomena (<20 s). *iF*-labelled arrows show an internal fusion process of the membranes and possibly combined with expulsion of matter out of the vesicle. *P*-labelled arrows refer to poration phenomena. Pores are stable in time and are used as "gates" for the inner vesicles (or tubes) to be expulsed (*E*-labelled arrows). Additional images (**Figure S 12**) illustrate further the membrane rearrangement and exvagination mechanism. Complementary scattering experiments (**Figure 7c** for MLVs and **Figure S 9a,b** for SUVs) show that the SAXS profile of mixed DOPC/G-C18:1 (0.6:1 molar ratio) at pH 8 can only be modeled with a thicker membrane (4.8 nm, **Figure 3b**) compared to DOPC/G-C18:1 at pH 5.5 (T= 4.0 nm), and rather in the same order as for DOPC MLVs, or SUVs (T= 4.8 nm).

Overall, changing the shape of G-C18:1 by increasing pH (cylinder to inverted cone) generates simple GUVs again through a mechanism of membrane disruption, hole formation and loss of cytoplasmic content. Considering the micellar morphology of G-C18:1 at alkaline pH, we performed a control experiment in which G-C18:1 micelles are added to a suspension of DOPC GUVs. **Figure S 13** shows that the micellar solution as such does not affect the morphology of the GUVs but it rather promotes budding (positive curvature) and small vesicles in close contact with the GUVs. The effect of deprotonated G-C18:1 seems to be in agreement with the effects observed for small amounts of deprotonated fatty acids or di-RLs, both characterized by an inverted conical shape. Deprotonated FAs are supposed to fuse to PC membranes through monomer diffusion,[52] they induce positive curvature,[14] they do not promote neither the $H_{II}$ phase nor fusion,[13] do not impact the vesicle size and trigger leaking.[52] Similar effects were reported for di-RLs: they generate defects and pores[30] favoring leaking[28,31] and facilitate budding and generation of daughter vesicles[30,32] in close contact with the mother vesicle, as observed here in **Figure S 13**. To note that high amounts of deprotonated FAs (e.g., oleic acid:Egg PC= 19:2 molar ratio) have a solubilizing effect (micellization) on egg PC.[12]

The effects of the deprotonated G-C18:1 on the mechanics of DOPC membrane is also demonstrated by the lower scaled bending rigidity ($\frac{k}{k_b T}$ = 0.4 ± 0.1) when compared to DOPC ($\frac{k}{k_b T}$ = 10.2 ± 0.5). If the contribution of micellar diffusion to such low values cannot be entirely excluded, we can reasonably rule it out using two arguments: the bending rigidity of a control sample composed of DOPC GUVs and G-C18:1 micelles is higher ($\frac{k}{k_b T}$ = 2.1 ± 0.1, **Figure 5**) and the SANS signal of DOPC/G-C18:1 at pH 10 is typical of a membrane (**Figure S 10c**, low-q slope is -2). Finally, lower bending rigidity of PC membranes is expected when containing charged amphiphiles of inverted cone shape.[53]

Acidification of the medium fully protonates G-C18:1. If it's shape remains cylindrical, low pH induces a transition from a vesicle suspension to a lamellar precipitate (**Figure 1**).[38,51] Full protonation of G-C18:1 after its inclusion in DOPC GUVs generates a two-step process (**Figure 7, Video S 5, Figure S 14**). Within the first 100 s (**Figure S 14a**), the entire content (vesicles, tubes) inside the GUVs and observed after 1 h equilibration (**Figure 7a)** diffuses towards and aggregates at the GUV membrane palisade. Between 100 s (**Figure S 14**) and 600 s (**Figure 7b**), spontaneous phase separation occurs within the membrane (**Figure 7b**), while vesicles aggregate. SAXS data recorded at pH 3 (**Figure 7c, Figure S 9c,** also displaying control DOPC and G-C18:1 signals) agree well with the microscopy data. The minimum of the form factor oscillation is at the same position as compared to pH 5.5, thus excluding major changes in the membrane thickness. Fitting of the SAXS profiles indeed show a constant membrane thickness (T= 4.1 ± 0.4 nm at pH 3, T= 4.0 ± 0.4 nm at pH 5.5). The low-q scattering profiles at q< 0.8 $nm^{-1}$ changes from a double to a single slope, confirming that the complex morphologies previously identified at pH 5.5 evolve into a simpler membrane morphology. Two dim, but visible, diffraction peaks ((001), (002), arrows in **Figure 7**c, pH 3, $q_{(001)}$= 0.942 $nm^{-1}$, d= 6.67 nm) are superimposed to the SAXS signal, thus suggesting a small fraction of multilamellar domains, agreeing well with the denser membrane portions of increased fluorescence intensities in **Figure 7b.**

NSE experiments nicely demonstrate how full protonation of G-C18:1 in the DOPC/G-C18:1 system leads to a ten-fold increase of the bending rigidity ($\frac{k}{k_b T}$ = 96.0 ± 31.0, **Figure 5**), without adding any other lipid or sterols. This phenomenon, as well as its magnitude, cannot be explained by the oleic acid moiety, as oleic acid is known to have a mild effect on the bending rigidity of DOPC.[40,50] The magnitude of $k$ is similar to the rigidity found in $L_\beta$ phases or the effect of cholesterol reported previously for DOPC.[54–56] However, in the present case, large values of $k$ are most likely associated with membrane aggregation and multilamellar domains, as seen by microscopy and SAXS measurements (**Figure 7**).

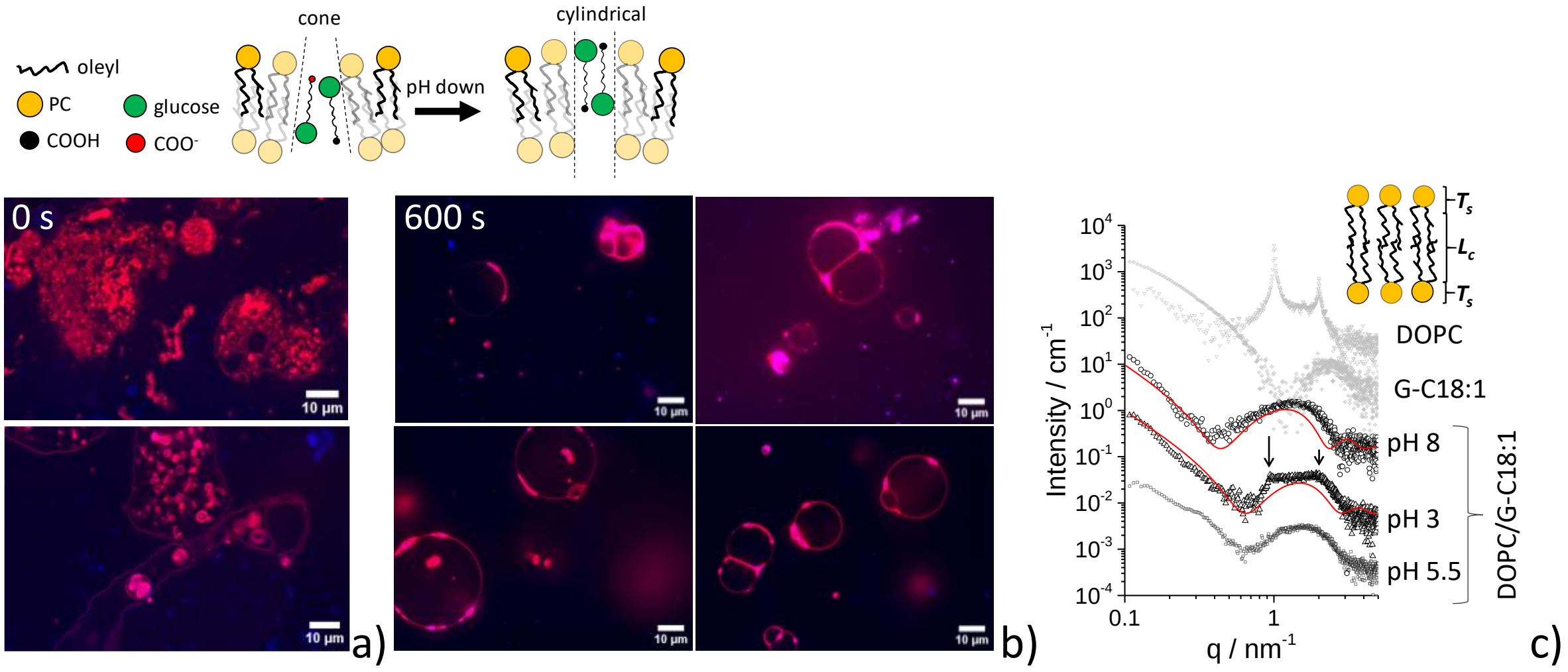

**Figure 7 – Reducing pH on DOPC-G-C18:1 intervesicle mixtures. Selected sites of a) equilibrated (1 h) (extracted from Video S 5) and b) acidified (1 M HCl) DOPC/G-C18:1 GUVs suspension. In b), images are recorded after 600 s of acid addition (Table 1 in the Experimental Methods section). c) Static SAXS experiments performed on DOPC MLVs and G-C18:1 vesicles. For sample DOPC/G-C18:1, [DOPC]:[G-C18:1]= 0.6 ($C_{DOPC}$= 2 mg/mL, $C_{G\text{-}C18:1}$= 1.8 mg/mL) at pH 5.5, pH 3 and pH 8. Controls at pH 5.5, reproduced from Figure 3a, are shown for sake of comparison. Samples at pH 3 and pH 8 are fitted with a head-tail lamellar form factor model (refer to the Supporting Information). Further details on sample preparation are given in Table S 1.**

## Mechanistic Considerations.

All of the results presented above can be reconciled in terms of a unifying mechanistic picture, such as developed below.

***First***, there are two limiting models of interactions between discrete amphiphilic mesostructures: molecular exchange and direct fusion. Confocal fluorescence microscopy (**Video S 1**) provides clear evidence for the existence of collision and coalescence-mediated interactions between the G-C18:1 and PC mesophases. This is further supported by the observed transfer of perylene to the DOPC membranes (**Figure S 4** to **Figure S 8**). The n-octanol/water partition coefficient, $K_{ow}$, of perylene is sufficiently high (log $K_{ow}$= 6.25)[57] that it is unlikely to permeate through molecular exchanges. The most likely scenario is consistent with the fusion of G-C18:1vesicles with DOPC vesicles. This mode of interactions at the mesoscopic (rather than molecular) scales is also consistent with our *in situ* SAXS measurements (**Figure 4**). The analyses of the SAXS data reveal a double exponential decay, which in turn suggests that the interactions between the two amphiphiles follows at least pseudo-first-order,[58] and possibly a second-order kinetics[58–60]. The coalescence-based mesoscopic interactions are uncommon for single-component lipid compartments but they are not surprising in this work. This could be explained by: (i) low bending rigidity ($< k_bT$) of G-C18:1 bilayers[61] ; (ii) double amphiphilic structure of G-C18:1 molecules, which generate packing defect in PC membranes in both straight and U-pin conformations; and (iii) thickness mismatch between G-C18:1 and PC bilayers. These concomitant factors can act synergistically to lower the activation energy barrier,[37,38] overcoming the entropic undulation repulsion between the vesicles [62] and driving fast (microsecond scale) fusion,[11] as hypothesized on **Figure 8**.

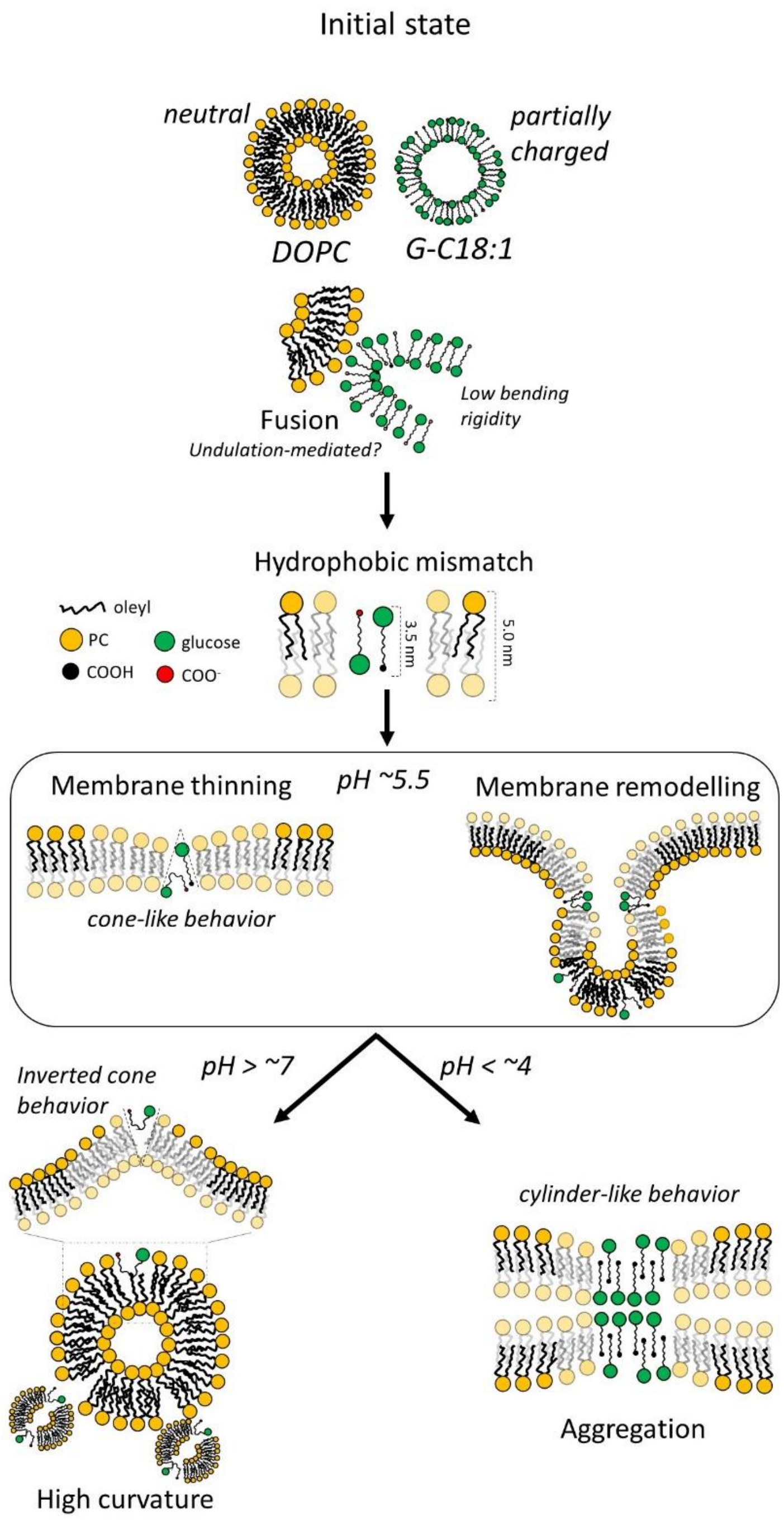

**Figure 8 – Proposed mechanism of the interaction of G-C18:1 and DOPC GUVs before and after pH change. The undulation-mediated fusion and conformation of G-C18:1 need further verification.**

***Second,*** we consider pH responsiveness of mixed DOPC/G-C18:1 vesicles. Our control experiments (**Figure S 3**) confirm previously reported findings[12,15] that DOPC vesicles are insensitive to pH changes. When all other factors, including osmotic balance, are maintained, pH changes alone do not perturb vesicular membranes. In agreement with the above, the low-q portion of the SAXS profiles (**Figure 7c, Figure S 9c**), the membrane thickness (**Figure 3b**) and bending rigidity (**Figure 5**, **Figure S 10**) also indicate no measurable changes.

In a marked departure, DOPC membranes containing G-C18:1 amphiphiles display a strong pH dependence. At low pH, in the absence of repulsive electrostatic forces (protonation of carboxyl groups), Van der Waals attraction and possibly sugar-sugar and sugar-COOH phenomena become the dominant sources of interactions, which could explain vesicle aggregation at pH< ~4, as illustrated in **Figure 8**. This hypothesis would be in correspondence with a previous report of SUV aggregation in aescin (a biosurfactant of triterpenoid saponin variety) doped DMPC/cholesterol vesicles, also driven by sugar-sugar interactions. At neutral/alkaline pH, by contrast, the vesicles behave quite differently. Here, deprotonation of G-C18:1 in the DOPC membrane can stabilize defects and strong local curvature, as shown for the pH> ~7 path in **Figure 8**. Mechanisms such as membrane poration, internal fusion, expulsion (**Figure 6b**, arrows labelled *P*, *iF* and *E*) but also fluctuations (**Figure S 11, Figure S 12),** and vesiculation (**Figure S 12, Figure S 13)** are possible and were all observed. These observations are qualitatively comparable to the effects of other inverted cone-shaped amphiphiles (like sodium oleate, rhamnolipids, or lysophosphatidylcholine) on PC membranes.

The relationship between the cone/inverted-cone effect and the molecular conformation of G-C18:1 is unverified and the molecular scale interpretation (bent vs. linear) depicted in **Figure 8** remains our working hypothesis. It is not unreasonable that G-C18:1 adopts both linear and bent configuration, as supposed for sophorolipids using molecular modelling arguments[63] and theorized for bolaamphiphiles in general.[16] Full deprotonation of G-C18:1 at pH> ~7 probably favors a bent configuration (**Figure 8**), assuming an effective cone shape, which stabilizes side defects and drives positive curvature. Full protonation at acidic pH, on the other hand, may favor the linear configuration of cylindrical shape and certainly eliminate repulsive electrostatic interactions, thus favoring Van der Waals attractive forces and sugar-sugar/sugar-COOH interactions (pH< ~4 in **Figure 8)**, with consequences on aggregation and increase in rigidity. Recalling that the pKa of G-C18:1 is 5.69,[34] both bent and linear conformations can exist in the intermediate pH range close to the pKa (4-7), with a resulting cone-like effect (pH< ~5.5 in **Figure 8**). This proposition also remains unverified and certainly requires further work.

## Conclusions

Using a combination of time-resolved confocal fluorescence microscopy, *in situ* small angle X-ray and neutron scattering (SAXS, SANS), and neutron spin echo (NSE) measurements, we have studied the interaction between nanoscopic vesicles of a bolaamphiphile, a glucolipid consisting of a single glucose headgroup and a C18:1 (oleyl) fatty acid tail (G-C18:1), with the vesicular membranes of an essentially cylindrical, fluid-phase phospholipid (dioleoylphosphatidylcholine, DOPC) in a pH-dependent manner. Our results reveal the two mesophases interact spontaneously and readily at all pH values. In all of the cases, the interaction gives rise to large-scale morphological remodeling. Under neutral and acidic conditions, when the bolaamphiphile is essentially cylindrical, vesicles fuse with one another, producing discrete vesicle aggregates, tubulation or vesosomes. Under basic pH, by contrast, when the carboxylic acid is deprotonated and the molecule is inverted-conical in shape, the bolaamphiphile causes phospholipid membranes to undergo large-scale topological changes. Here, phospholipid membranes exhibit poration, budding, and vesiculation. This pH-dependent, environmentally sensitive membrane remodeling illustrates how local, molecular-level packing perturbations can translate into global system-level morphological changes, enabling membranes to acquire environmental sensitivity and real-time adaptability within the seconds time-scale. Our results support the notion that molecular fluxes – which add (or remove) amphiphilic molecules to biological membranes – can endow *de novo* functionalities (e.g., pH sensitivity) and influence global morphologies of cell-sized membranous structures. They also suggest that bolaamphiphiles can be used to engineer environmental sensitivity in synthetic vesicles and model protocells.

## Experimental Methods

### Chemicals

18:1 (Δ9-*cis*) PC, 1,2-dioleoyl-sn-glycero-3-phosphocholine (DOPC) is purchased from Avanti Lipids. G-C18:1 is purchased from Amphistar (Gent, Belgium), batch No. APS F06/ F07, Inv96/98/99 and used as such. The monounsaturated glucolipid G-C18:1 (Mw = 460 $g.mol^{-1}$) contains a β-D-glucose unit covalently linked to oleic acid (Figure 1 in the main text). The molecule is obtained by fermentation from the yeast *Starmerella bombicola* ΔugtB1 according to the protocol given before.[64] Although the known IUPAC name for this compound (17-O-(β-D-glucopyranosyl)-octadecenoic acid[64,65] does not specify the nature of the stereoisomer (L or D), its molecular similarity with the sophorose lipids named 17-L-[(2′-O-β-glucopyranosyl-β-D-glucopyranosyl)-oxy]-9-octadecenoic acids[66] suggesting that the L-form is dominant.
According to the specification sheet provided by the producer, the batch (99.4 % dry matter) was composed of 99.5 % G-C18:1, according to HPLC-ELSD chromatography data. NMR analysis of the same compound (different batch) was performed elsewhere.[38] HCl (37%) and NaOH pellets are purchased from Sigma Aldrich. Perylene is purchased from Sigma Aldrich. 18:1 *Liss* Rhod *PE*, 1,2-dioleoyl-sn-glycero-3-phosphoethanolamine-N-(lissamine rhodamine

B sulfonyl) (ammonium salt) (LISS-PE) is purchased from Avanti Lipids. 4′,6-diamidino-2-phenylindole (DAPI) is purchased from Sigma Aldrich.

**Label solutions**

Red label is prepared from a rhodamine-based ($\lambda_{ex}$= 561 nm) dye: 1 mg/mL of LISS-PE is prepared in chloroform. Blue label is prepared from perylene or DAPI ($\lambda_{ex}$= 405 nm) dyes. For perylene, a saturated solution is prepared at 5 mg/mL in MeOH. For DAPI, a 1 mg/mL solution is also prepared in MeOH. Green label is prepared from fluorescein ($\lambda_{ex}$= 488 nm): 1 mg/mL carboxyfluorescein is prepared in water at pH 5.5.

**Electroformation of GUVs**

*DOPC solution*: 40 µL of a 25 mg/mL (solvent: $CHCl_3$) DOPC solution are mixed with 17 µL of a 1 mg/mL LISS-PE ($\lambda_{ex}$= 561 nm, solvent: $CHCl_3$), to which 444 µL of $CHCl_3$ are added. LISS-PE represents 1 % molar ratio of DOPC. The concentration of the DOPC solution used for electroformation is 2.6 mM (2 mg/mL).

*G-C18:1 solution*: 40 µL of a 15 mg/mL (solvent: $CHCl_3$/MeOH at 70/30 v/v) solution of G-C18:1 is mixed with a 17 µL of 1 mg/mL LISS-PE ($\lambda_{ex}$= 561 nm, solvent: $CHCl_3$), to which 444 µL of $CHCl_3$/MeOH solvent mixture at 70/30 v/v is added. LISS-PE represents 1 % molar ratio of G-C18:1. The concentration of the G-C18:1 solution used for electroformation is 2.6 mM.

*Sample preparation*: 15 µL of either the DOPC or the G-C18:1 solutions are spread on the conductive side of an ITO slide as homogeneously as possible. The slides are allowed to dry at room temperature under vacuum during about 1 h. After drying, a rubber o-ring is gently adapted on the conductive side of the slide, just above the dried sample. Silicon vacuum grease is used to seal the o-ring to the slide. The o-ring cavity is filled with an aqueous solution of milliQ-grade water containing 0.1 M sucrose. Silicon vacuum grease is also gently placed on the top side of the o-ring, onto which the second ITO slide is eventually placed. The sample-containing conductive side faces the sucrose solution. Extra water squeezing out of the sandwich is dried and the slide/o-ring/slide sandwich itself is eventually clipped together. The sandwich is connected to the electric function generator by means of the electrodes, each one clipped at each side of the ITO slides.

*Electric function generator*: the following parameters were found to be optimal to produce GUVs for both the DOPC and G-C18:1 samples. However, the actual time can be quite flexible and span between 40 min to 120 min, especially for DOPC. Use of square wave in the second cycle seemed to the authors to improve the GUVs production of G-C18:1, although its use should not be taken as mandatory. Cycle 1: 40 min using a sine wave shape at a frequency of 10 Hz and peak-to-peak amplitude of 3 V. Cycle 2: 20 min using a square wave shape at a frequency of 10 Hz and peak-to-peak amplitude of 2.2 V. There is no time gap between cycle 1 and cycle 2.

**Vesicle suspension of G-C18:1**

Vesicle preparation for G-C18:1 is adapted from previous work.[38] A 5 mg/mL solution of G-C18:1 is prepared in milliQ-grade water containing sucrose at concentration of 0.1 M for equimolarity reasons. The sample is sonicated in a sonicator bath until the powder is finely dispersed. pH is then adjusted to 5.5 using few µL amounts of 0.5 M and 0.1 M NaOH solution. The solution is vortexed between each aliquot addition. If needed, sonication and gentle warming (<40°C) can be used in support to vortexing. This process stops when pH is stable and the solution is homogenously turbid with no sedimentation occurring and no sign of large aggregates. At this stage, vesicles are formed. This solution can be used as such, or labeled with perylene. In the latter, 1 µL of the saturated perylene ($\lambda_{ex}$= 405 nm) solution in MeOH is added to 1 mL of the G-C18:1 vesicle solution (water, pH 5.5). The vesicle solution can eventually be diluted by a factor 5 (concentration of G-C18:1 is 1 mg/mL) or 10 (concentration of G-C18:1 is 0.5 mg/mL) using a 0.1 M sucrose solution at pH 5.5.

**Sample preparation and study for confocal microscopy**

All confocal microscopy experiments involve the use of DOPC GUVs suspensions prepared by electroformation[67] and G-C18:1 vesicle suspensions prepared by direct dispersion in water (see above). Samples are prepared in a 96-well plate equipped with a glass bottom and settled in the appropriate sample holder under the microscope. Samples are prepared immediately before each observation and experiments are repeated at least three times. Samples are numbered from (1) to (6). All solutions are systematically osmotically balanced to prevent mechanical stress of the membranes.

*(1) DOPC(-LISS-PE) GUVs control*: typically, 96 µL of a 0.1 M glucose solution at pH 5.5 is pipetted into a single well, to which 1 µL of electroformed Liss-PE labelled DOPC GUVs in 0.1 M sucrose is added. Due to the difference in density between sucrose (inside GUVs) and glucose (medium), GUVs sediment at the bottom of the well within a minute time span. The osmotically balanced sucrose (lumen of GUVs) and glucose (medium) media prevent mechanical stress of the GUVs.[68] Selected samples were imaged after equilibrating *(2)* at room temperature during 1 h. This sample is observed as such using a laser at $\lambda_{ex}$= 561 nm.

*(2) G-C18:1(-perylene) vesicles in DOPC GUVs pH 5.5*: typically, to *(1)* (*DOPC GUVs control*), one adds 1 µL of a 5 mg/mL G-C18:1 vesicle solution (pH 5.5, 0.1 M sucrose), which can be added as such or containing perylene. *In situ* observations using the confocal microscope start before adding the G-C18:1 vesicle solution using the laser wavelengths at $\lambda_{ex}$= 561 nm and $\lambda_{ex}$= 405 nm. Selected samples were imaged after equilibrating *(2)* at room temperature (23 °C ± 1°C during 1 h). This sample is observed as such using a combination of lasers at $\lambda_{ex}$= 561 nm and $\lambda_{ex}$= 405 nm.

*(3) G-C18:1(unlabeled) vesicles in DOPC GUVs using DAPI*: in a variant experiment, having the goal of quantifying the insertion of G-C18:1 in the DOPC GUVs, unlabeled (no perylene) G-C18:1 vesicles are added to *(1)* (*DOPC GUVs control*) to which 1 µL of a 1 mg/mL DAPI solution was previously added. Confocal microscopy imaging starts before the G-C18:1 solution is added. This sample is observed as such using a combination of lasers at $\lambda_{ex}$= 561 nm and $\lambda_{ex}$= 405 nm.

*(4) NaOH in G-C18:1 vesicle / DOPC GUVs pH 5.5*. 1 µL of 0.1 M NaOH solution is added to *(2)* (*G-C18:1(-perylene) vesicles in DOPC GUVs*), for a concentration of NaOH of 1 mM and pH raising to about 9. Observation at confocal microscope starts on *(2)* before adding the source of NaOH. This sample is observed as such using a combination of lasers at $\lambda_{ex}$= 561 nm and $\lambda_{ex}$= 405 nm.

*(5) G-C18:1 micelles in DOPC GUVs control*: as additional control, a micellar solution is added to DOPC GUVs. Micelles are prepared from G-C18:1 at pH 8 according to Ref. [38] by preparing a 5 mg/mL G-C18:1 solution in water at pH 8, where the compound becomes soluble and the solution clear. Labelling is performed by adding 1 µL of the perylene solution (5 mg/mL in MeOH) to 1 mL of the G-C18:1 micellar solution. Eventually, 1 µL of the micellar G-C18:1 solution is added to *(1)* (*DOPC GUVs control*), to which 2 µL of a 0.1 M NaOH solution were preliminarily added to increase pH to basic. Confocal microscope imaging starts before adding the micellar solution to *(1)*. This sample is observed as such using a combination of lasers at $\lambda_{ex}$= 561 nm and $\lambda_{ex}$= 405 nm.

*(6) HCl in G-C18:1 vesicle / DOPC GUVs*: 1 µL of 1 M HCl solution is added to *(2)* (*G-C18:1(-perylene) vesicles in DOPC GUVs*), for a concentration of HCl of 10 mM and pH lowering to about 3.5. Observation at confocal microscope starts on *(2)* before adding the source of HCl. This sample is observed as such using a combination of lasers at $\lambda_{ex}$= 561 nm and $\lambda_{ex}$= 405 nm.

**Table 1– Details of experimental conditions associated to the confocal images experiments presented in the main text**

| Figure | Experimental conditions |
|---|---|
| Figure 2 | V= 96 µL, $V_{DOPC\ GUVs}$= 1 µL, $V_{G\text{-}C18:1}$= 1 µL, $C_{G\text{-}C18:1}$= 5 mg/mL or 0.5 mg/mL, pH= 5.5, T= 23 ±1 °C, frequency of acquisition, f= 20 Hz. |
| Figure 6 | V= 96 µL, $V_{DOPC\ GUVs}$= 1 µL, $V_{G\text{-}C18:1}$= 1 µL, $C_{G\text{-}C18:1}$= 5 mg/mL, $V_{NaOH}$ = 1 uL, [NaOH]= 0.1 M, T= 23 ±1 °C, frequency of images acquisition, f= 20 Hz. |
| Figure 7a,b | V= 96 µL, $V_{DOPC\ GUVs}$= 1 µL, $V_{G\text{-}C18:1}$= 1 µL, $C_{G\text{-}C18:1}$= 5 mg/mL, $V_{HCl}$= 2 µL, [HCl]= 1 M; T= 23 ±1 °C, frequency of images acquisition, f= 20 Hz. |

## Confocal microscopy

Marianas 3i Spinning disk confocal microscope using a 96-wells plate support and a 60x oil objective (N. A. 1.4). The microscope uses a EMCCD photometrics cascade camera with 50x gain. Unless otherwise stated, images are acquired every 200 ms, that is a frequency of 5 images per second.

## Small angle X-ray scattering (SAXS)

SAXS experiments were performed at room temperature on the BM26 Beamline at ESRF Synchrotron (Grenoble, France). The BM26 beamline was used with an energy of E = 12 KeV, a fixed sample-to-detector Eiger X 4M distance of 1.95 m and a borosilicate capillary of 2 mm in diameter. q is the wave vector ($q = 4\pi/\lambda \sin(\theta)$, 2θ correspond to the scattering angle and λ the wavelength). The q-range is calibrated between $\sim 0.1 < q / nm^{-1} < \sim 6$ ; raw data obtained on the 2D detector are integrated azimuthally using the in-house software provided at the beamline and thus to obtain the typical scattered intensity I(q) profile. Absolute intensity units are determined by measuring the scattering signal of water ($I_{q=0} = 0.0163\ cm^{-1}$).

The conditions of sample preparation for SAXS experiments as well as the method to fit the SAXS data are given in the Supporting Information available online.

## Neutron scattering experiments.

Small angle neutron scattering (SANS) and neutron spin echo (NSE) experiments were performed during the same beamtime (proposal number 9-13-1102[69]) at the Institut Laue-Langevin (ILL) in Grenoble (France) on the same set of samples. The specific conditions employed to prepare samples for SANS and NSE are given in the Supporting Information available online.

*Small angle neutron scattering***.** SANS experiments were performed on the instrument D22 beamline. Experiments were carried out at two simultaneous detector-to-sample distances (1.4 and 17.6 m) and two neutron wavelengths, λ= 6 Å (high-q range) and λ= 11.5 Å (low-q range) to cover a full $q$-range of 0.0014 $Å^{-1}$ < $q$ < 0.63 $Å^{-1}$, with $q = \frac{4\pi}{\lambda} sin\vartheta$ , where $2\vartheta$ is the scattering angle between the incident and the scattered neutron beam. The rear detector (17.6 m) was centered on and perpendicular to the direct beam while the front detector (1.4 m) was tilted at 20 ° from the sample to provide a continuous angular coverage of the detection. A semi-transparent attenuator (B10-enriched B4C) was used to measure simultaneously the transmission. The beam was collimated over 17.6 m, with a circular source aperture of 30 mm diameter and a sample aperture of 7 (horizontal) x 10 (vertical) $mm^2$. Samples were contained in a quartz cuvette of 1 mm pathway (110-QS and 120-QS, Hellma GmbH & Co. KG, Müllheim, Germany), installed in a thermalized copper sample changer (25°C). The data were processed with the program Grasp V.10.24d,[70] taking into account the detector background (measurement of sintered $^{10}B_4C$), empty container, sample thickness, transmission as measured within the direct beam, parallax, normalizing with monitor and correcting with a flat field, using experimental beam centers for both detectors. Absolute scale was obtained from the measurement of the empty beam flux ($1.05 \cdot 10^6$ neutrons per second), as measured on the rear detector with a chopper having a known attenuation coefficient. The data were binned according to q-resolution with 5 bins per standard deviation. This resolution was calculated from the direct beam profile. The samples were prepared in deuterated water. The spectrometer resolution is Δλ/λ= 10%. Acquisition time was set at 5 min at high-q and 10 min at low-q. Data were azimuthally averaged to yield the typical 1D intensity distribution $I(q)$. All samples studied by SANS were prepared in $D_2O$ (see Supporting Information available online for more information on sample preparation for SANS).

Neutron spin echo (NSE) experiments. NSE[71] measurements have been performed at the instrument IN15.[72] Four different wavelengths (λ) have been used, namely 13.5, 12, 10 and 8 Å allowing to reach maximum Fourier times $t = \frac{\gamma_N J_i m_N^2}{2\pi h^2} \lambda^3$ of 477, 335, 194 and 99 ns, respectively, where $\gamma_N$ and $m_N$ are the neutron's gyromagnetic ratio and mass, $J_i$ is the instrument's field integral and $h$ is Planck's constant. At the same time, we are covering a $q$-range from 0.020 to 0.128 $Å^{-1}$, where $q = \frac{4\pi}{\lambda} sin\left(\frac{\theta}{2}\right)$ is the modulus of the scattering vector, with scattering angle $\theta$, here chosen to be 8.5° (λ= 8 Å), 7.5° (λ= 10 Å), 6.2° (λ= 12 Å) and 3.5° ($\lambda$= 13 Å). The data were corrected for resolution effects using graphite and the scattering from the aqueous background was subtracted. The data analysis protocol adopted to analyze NSE data is given in the Supporting information available online.

## Supporting Information

Supporting Information files are available online: Additional experimental method section for the preparation of samples for scattering experiments, method of fitting SAXS data and treating NSE data (Page S2 to S7), legends of Video S 0 to Video S 5 (videos are available online) and supplementary figures, Figure S 1 (chemical structures of the chromatophores), Figure S 2 (microscopy of controls vesicles), Figure S 3 (confocal microscopy of DOPC GUV controls), Figure S 4 to Figure S 7 (confocal microscopy of DOPC GUVs mixed with G-C18:1 vesicles), Figure S 8 (confocal microscopy of the effect of G-C18:1 vesicles on GUVs of mixed lipid composition), Figure S 9 (SAXS of mixtures of DOPC SUVs and G-C181 vesicles), Figure S 10 (small angle neutron scattering and neutron spin echo spectroscopy of DOPC SUVs mixed with G-C181 vesicles), Figure S 11 to Figure S 12 (confocal microscopy showing the effect of NaOH on mixtures of DOPC GUVs and G-C181 vesicles), Figure S 13 (confocal microscopy showing the effect of G-C18:1 micelles on DOPC GUVs).

Video S 0 and Video S 1 present two experimental replicates showing the time-resolved in situ evolution of DOPC GUVs suspension to which an aliquot of G-C18:1 (5 mg/mL) vesicles suspension is added. Video S 2 and Video S 3 present two experimental replicates showing the time-resolved in situ evolution of DOPC GUVs suspension to which an aliquot of G-C18:1 (0.5 mg/mL) vesicles suspension is added. Video S 4 presemts a time-resolved in situ evolution of DOPC/G-C18:1 GUVs suspension (equilibration time: 1 h) to which a 0.1 M NaOH solution is added. Video S 5 présents a time-resolved in situ evolution of DOPC/G-C18:1 GUVs suspension (equilibration time: 1 h) to which a 1 M HCl solution is added.

## Acknowledgements

The author gratefully thanks Chris Carnahan and Thomas Wilkop (UC Davis, CA, USA) for their assistance on the GUV preparation and confocal microscopy, respectively. The France-USA Fulbright Commission (award N° PS00342290) is kindly acknowledged for supporting this research. Additional support for the work at UC Davis was obtained through a grant from the National Science Foundation (Award #2342436, Biomaterials program, division of materials research). The Agence Nationale de la Recherche (ANR), funding ANR-22-CE43-0017, is also acknowledged for funding this work. The European Synchrotron, ESRF, is kindly acknowledged for granting access the BM29 beamline. Institut Laue-Langevin, ILL, (Grenoble, France) is kindly acknowledged for financial support related to the proposal 9-13-1102.

# Supporting Information

## Driving a stimuli-responsive wedge in the packing of phospholipid membranes using bolaamphiphile intercalants

Niki Baccile,[a,*] Archan Vyas,[b] Ramanujam Ramanujam,[a] Daniel Hermida-Merino,[c] Ingo Hoffmann,[d] Lionel Porcar,[d] Atul N. Parikh[b,e,*]

[a] Sorbonne Université, Centre National de la Recherche Scientifique, Laboratoire de Chimie de la Matière Condensée de Paris, LCMCP, F-75005 Paris, France

[b] Departments of Biomedical Engineering and Chemical Engineering, University of California, Davis, California 95616, USA

[c] Departamento de Física Aplicada, CINBIO, Universidade de Vigo, Campus Lagoas-Marcosende, Vigo, 36310, Spain

[d] Institut Laue-Langevin, 38042 Grenoble, France

e School of Materials Science & Engineering, Singapore Centre for Environmental Life Sciences & Engineering, and Institute for the Digital Molecular Analytics & Science, Nanyang Technological University, Singapore

*** Corresponding author:**

Dr. Niki Baccile

niki.baccile@sorbonne-universite.fr

Prof. Atul Parikh

anparikh@ucdavis.edu

## Experimental section

### SAXS experiments: conditions of sample preparation

Two types of experiments are performed, static and *in situ* with time resolved acquisition. In both cases, a given volume of the G-C18:1 vesicle suspension (C= 20 mg/mL, pH 5.5, prepared by bath sonication 10 min at 40°C) is added to a DOPC liposomal solution (C= 2 mg/mL, pH 5.5), of which the detailed preparation procedure is detailed below. In the static experiments, the G-C18:1 and DOPC suspensions are mixed, inserted by manual injection in the capillary and analyzed after about 20 min of equilibration. When pH is changed, an additional equilibration time of 10 min is added after pH variation. More experimental details are given in Table S 1, below.

In the time resolved *in situ* experiment, aliquots of a given volume of the G-C18:1 suspension are added at once (1 s) into the DOPC suspension with intervals of 5 min over a total time of 20 min (5 additions in total). The reason behind this experimental approach is explained by the objective of finding the minimum amount of G-C8:1 having effects on DOPC GUVs. The SAXS signal is acquired continuously over 20 min without intervals and with an acquisition time per spectrum of 5 s. To improve the signal-to-noise ratio, the SAXS profiles published in this work are averaged over 12 spectra, corresponding to 60 s of total acquisition, after verification that all spectra within this time range are superimposable. All details about the sample preparation conditions are given in Table S 1. The *in situ* experiment is performed following a well-known setup in our Team[1] and involving the use of a peristaltic pump continuously pumping the sample through the capillary. Addition of the G-C18:1 suspension is performed from the experimental hutch with a remotely-controlled (volume and rate) push syringe (New Era, NE1000, freeware: Pumpterm).

Given the low concentration of the liposomal solution prepared by electroformation, all samples used for SAXS and SANS experiments were prepared by adapting the classical thin film hydration and extrusion method.[2,3] 40 mg of DOPC were dissolved by sonication (bath) in chloroform, which was vacuum pumped in a 50 mL round bottom flask using a rotavapor (45°C, 180 mbar, 80 rpm), whereas the classical spinning helps distributing the DOPC homogeneously on the bottom of the flask. The flask was then allowed to dry 24 h under high vacuum at 40°C to remove extra chloroform. 1 mL of milliQ water was added to the flask, which was sonicated at about 40°C twice during 20 min. The solution was diluted by a factor two, to reach 20 mg/mL (mother solution) and sonicated again twice at 40°C. From the mother solution, several step solutions at 2 mg/mL or 1 mg/mL were prepared by diluting with milliQ water, the pH naturally being at about 5.5. These solutions are freeze thawed twice (-18°C; +40°C) and eventually either further sonicated (bath) 45 min at 40°C to yield MLVs, or extruded using an Avanti mini extruder using a pore diameter of 0.1 µm (15 passes) at room temperature, to yield single unilamellar vesicles (SUVs). The

**Table S 1 – Details of preparation conditions associated to the DOPC, G-C18:1 and DOPC/G-C18:1 samples studied under static and *in situ* condition by SAXS. The terms MLVs and SUVs is employed in Table S 1 refer to DOPC MLVs or SUVs suspensions, of which the method of preparation is given in the SAXS section of the Supporting Information.**

| Experiment and location in figures | Description of the samples |
|---|---|
| **Experiments on MLVs** | |
| Static SAXS on MLVs<br><br>Figure 3a | - Sample G-C18:1 → bath sonication, C= 5 mg/mL, pH 5.5. Acquisition time= 60 s<br><br>- Sample DOPC MLVs → gentle hydration, C= 2 mg/mL, pH 5.5. Acquisition time= 60 s<br><br>- Sample DOPC/G-C18:1 → Note: 150 µL of a G-C18:1 vesicle suspension (20 mg/mL) are added to 2 mL of a 2 mg/mL DOPC MLVs suspension, both at pH 5.5<br>Concentrations:<br>$C_{DOPC}$= 2 mg/mL, $V_{DOPC}$= 2 mL, Mw $_{DOPC}$= 786.1 g/mol, [DOPC]= 2.5 mM<br>$C_{G\text{-}C18:1}$= 1.4 mg/mL, Mw $_{G\text{-}C18:1}$= 460.0 g/mol, [G-C18:1]= 3.0 mM<br>[DOPC]:[G-C18:1]= 0.8<br><br>Each SAXS spectrum is averaged over 12 spectra, corresponding to 60 s of acquisition time (acquisition time of each spectrum is 5 s). |
| Time resolved in situ SAXS on MLVs<br><br>Figure 4 | Five aliquots of 50 µL of G-C18:1 vesicle suspension (20 mg/mL) are added to 2 mL of a 2 mg/mL DOPC MLVs suspension every 5 min for a total experimental time of 20 min.<br><br>Concentrations and molar ratios:<br>$C_{DOPC}$= 2 mg/mL, $V_{DOPC}$= 2 mL, Mw $_{DOPC}$= 786.1 g/mol, [DOPC]= 2.5 mM, pH 5.5<br>$C_{G\text{-}C18:1}$= from 0.5 to 2.2 mg/mL, Mw $_{G\text{-}C18:1}$= 460.0 g/mol, [G-C18:1]= from 1.1 to 4.8 mM; pH 5.5<br>Evolution of molar ratio in time:<br>time: 0 min, [DOPC]:[G-C18:1]= 2.3<br>time: 5 min, [DOPC]:[G-C18:1]= 1.2<br>time: 10 min, [DOPC]:[G-C18:1]= 0.8<br>time: 15 min, [DOPC]:[G-C18:1]= 0.6<br>time: 20 min, [DOPC]:[G-C18:1]= 0.5<br><br>Each SAXS spectrum is averaged over 6 spectra, corresponding to 30 s of acquisition time (acquisition time of each spectrum is 5 s). |
| Static SAXS for pH change experiments on MLVs<br><br>Figure 7c | - Sample G-C18:1 → bath sonication, C= 5 mg/mL, pH 5.5. Acquisition time= 60 s<br><br>- Sample DOPC MLVs → gentle hydration, C= 2 mg/mL, pH 5.5. Acquisition time= 60 s<br><br>- Sample DOPC/G-C18:1 → 100 µL of a G-C18:1 vesicle suspension (20 mg/mL) are added to 1 mL of a 2 mg/mL DOPC MLVs suspension, both at pH 5.5. Acquisition time= 60 s<br>Concentrations:<br>$C_{DOPC}$= 2 mg/mL, $V_{DOPC}$= 1 mL, Mw $_{DOPC}$= 786.1 g/mol, [DOPC]= 2.5 mM |
| | $C_{G\text{-}C18:1}$= 1.8 mg/mL, Mw $_{G\text{-}C18:1}$= 460.0 g/mol, [G-C18:1]= 3.9 mM<br>[DOPC]:[G-C18:1]= 0.6<br><br>- Sample DOPC/G-C18:1 at pH 5.5 is split in half and pH is raised to 8 or lowered to 3. Samples are allowed an equilibration of 10 min after changing pH. Acquisition time= 60 s |
| **Experiments on SUVs** | |
| Static SAXS on SUVs<br><br>Figure S 9a | - Sample G-C18:1 $\rightarrow$ bath sonication, C= 5 mg/mL, pH 5.5. Acquisition time= 60 s<br><br>- Sample DOPC SUVs $\rightarrow$ gentle hydration, C= 2 mg/mL, pH 5.5, freeze thawed twice and sonicated 40°C twice for 10 min, filtered with 0.1 µm, 15 passes. Acquisition time= 60 s<br><br>- Sample DOPC/G-C18:1 $\rightarrow$ Note: 100 µL of a G-C18:1 vesicle suspension (20 mg/mL) are added to 2 mL of a 2 mg/mL DOPC SUVs suspension, both at pH 5.5<br><br>Concentrations:<br>$C_{DOPC}$= 2 mg/mL, $V_{DOPC}$= 2 mL, Mw $_{DOPC}$= 786.1 g/mol, [DOPC]= 2.5 mM<br>$C_{G\text{-}C18:1}$= 0.95 mg/mL, Mw $_{G\text{-}C18:1}$= 460.0 g/mol, [G-C18:1]= 2.0 mM<br>[DOPC]:[G-C18:1]= 1.2<br><br>Each SAXS spectrum is averaged over 12 spectra, corresponding to 60 s of acquisition time (acquisition time of each spectrum is 5 s). |
| Time resolved *in situ* SAXS on SUVs<br><br>Figure S 9b | Five aliquots of 20 µL of G-C18:1 vesicle suspension (20 mg/mL, sonicated 10 min at pH 5.6 and T= 40°C) are added to 2 mL of a 2 mg/mL DOPC SUVs (freeze thawed twice and sonicated 40°C twice for 10 min, filtered with 0.1 µm, 15 passes) suspension every 5 min for a total experimental time of 20 min.<br><br>Concentrations and molar ratios:<br>$C_{DOPC}$= 2 mg/mL, $V_{DOPC}$= 2 mL, Mw $_{DOPC}$= 786.1 g/mol, [DOPC]= 2.5 mM, pH 5.5<br>$C_{G\text{-}C18:1}$= from 0.2 to 0.95 mg/mL, Mw $_{G\text{-}C18:1}$= 460.0 g/mol, [G-C18:1]= from 0.43 to 2.1 mM; pH 5.5<br>Evolution of molar ratio in time:<br>time: 0 min, [DOPC]:[G-C18:1]= 5.9<br>time: 5 min, [DOPC]:[G-C18:1]= 2.9<br>time: 10 min, [DOPC]:[G-C18:1]= 2.0<br>time: 15 min, [DOPC]:[G-C18:1]= 1.5<br>time: 20 min, [DOPC]:[G-C18:1]= 1.2<br><br>Each SAXS spectrum is averaged over 6 spectra, corresponding to 30 s of acquisition time (acquisition time of each spectrum is 5 s). |
| Static SAXS for pH change experiments on SUVs<br><br>Figure S 9c | - Sample G-C18:1 $\rightarrow$ bath sonication, C= 5 mg/mL, pH 5.5. Acquisition time= 60 s<br><br>- Sample DOPC SUVs $\rightarrow$ gentle hydration, C= 2 mg/mL, pH 5.5. Acquisition time= 60 s<br><br>- Sample DOPC/G-C18:1 $\rightarrow$ 100 µL of a G-C18:1 vesicle suspension (20 mg/mL) are added to 1 mL of a 2 mg/mL DOPC SUVs suspension (extrusion 0.1 µm, 15 passes), both at pH 5.5. Acquisition time= 60 s<br>Concentrations:<br>$C_{DOPC}$= 2 mg/mL, $V_{DOPC}$= 1 mL, Mw $_{DOPC}$= 786.1 g/mol, [DOPC]= 2.5 mM<br>$C_{G\text{-}C18:1}$= 1.8 mg/mL, Mw $_{G\text{-}C18:1}$= 460.0 g/mol, [G-C18:1]= 3.9 mM<br>[DOPC]:[G-C18:1]= 0.6<br><br>- Sample DOPC/G-C18:1 at pH 5.5 is split in half and pH is raised to 8 or lowered to 3. Samples are allowed an equilibration of 10 min after changing pH. Acquisition time= 60 s |

70

**SAXS experiments: Fitting methodology**

All SAXS profiles, except DOPC MLVs, were fitted with a head-tail lamellar form factor model (LamellarFFHGModel in the SasView 3.1.2 software[4]). The DOPC MLVs was fitted with a lyotropic lamellar phase form and structure factor model using the Nallet et al.[5] approach based on Caillé's model[6] (LamellarPSHGModel in the SasView 3.1.2 software[4]). In short, the lamellar model requires a polar hydrophilic shell thickness, $T_s$, and a core length, $L_c$, being the sum of twice the tail's length, $L$. The total membrane thickness is then, T= $2Ts + 2L$= $2T_s+L_c$, as shown in the sketch hereafter:

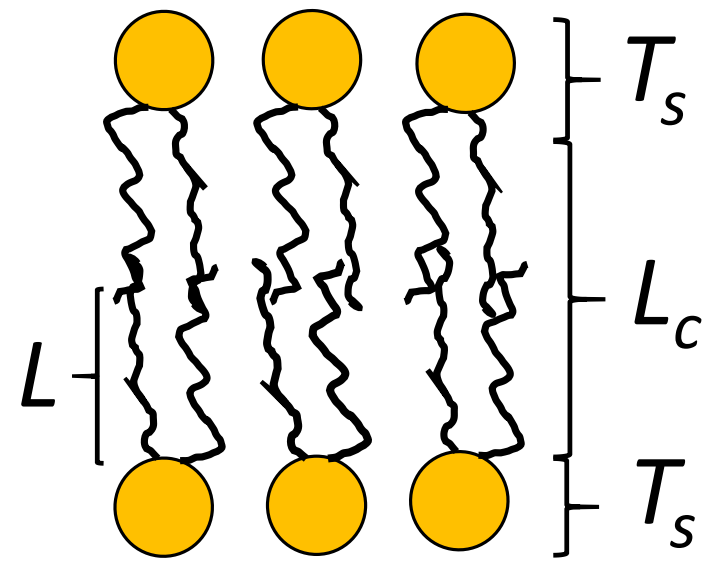

The model also requires the use of standard fitting parameters like the scattering length density (SLD), $\rho$, of the shell, core and solvent, noted $\rho_s$, $\rho_c$ and $\rho_{solvent}$, respectively, the scale and the background. The fitting process was adapted from previous work.[1,7] Here, we fix the scale value at 1, $\rho_c$= 8.4 x $10^{-4}$ $nm^{-2}$ and $\rho_{solvent}$= 9.4 x $10^{-4}$ $nm^{-2}$, while $\rho_s$ was allowed to vary within the limits of 10.0 and 13.0 x $10^{-6}$ $nm^{-2}$, a known range which accounts for a hydrated headgroup of the amphiphiles. The background is set at 0.0003 $cm^{-1}$, although allowed a ±0.0001 $cm^{-1}$ variation if needed. Eventually, only $T_s$, $\rho_s$ and $L$ are the free variables and all constrained within physically reasonable values, estimated from previous work.[1,7] The values found for $T_s$ and $L$ are given in Figure 3b in the main text, while $\rho_s$ was found to be between 11.6 x $10^{-4}$ $nm^{-2}$ and 12.0 x $10^{-4}$ $nm^{-2}$. In practice, the fit of the G-C18:1 signal was based on previous work.[1,7] For the other samples, the fitting parameters were optimized on the DOPC SUVs sample and then used as such to model the signal of the remaining samples. It was found that the only significant parameter affecting the model was $L$, $T_s$ and $\rho_s$ needing only minor adjustments. For the DOPC MLV system, the Caillé parameter, $\eta$, is 0.15 and the number of plates 100. Low values for $\eta$ are expected for lamellar phases with long-range order. Finally, all fitted values should be considered within a 10% error margin.

**Neutron scattering experiments: conditions for the sample preparation.**

The preparation method was essentially the same as for SAXS, exception made for the use of $D_2O$, DCl (0.5 M, 1 M and 5 M) and NaOD (0.5 M, 1 M and 5 M) instead of hydrogenous water. Use of deuterated solvents is critical to reduce incoherent scattering and to increase contrast. DCl and NaOD solutions were prepared from 35 wt% DCl and pellet NaOH, respectively. Differently than for SAXS, only SUVs were analyzed in neutron experiments and this is explained by the need of eliminating any trace of the (001) lamellar peaks, which can generate artefacts in the NSE signal due to the de Gennes narrowing effects.[8,9] Shortly, DOPC in $D_2O$ were prepared at C= 2 mg/mL following the gentle hydration procedure described for the SAXS experiments. The DOPC vesicle suspension was then systematically extruded using the Avanti mini extruder with a pore diameter of 0.1 µm (21 passes) and so to obtain SUVs only. Differently than in SAXS, extrusion (21 steps) was also performed on samples prepared from DOPC and mixed with G-C18:1, so to promote mixing between the two compounds and thus reducing the morphological complexity of the samples and which could lead to interpretation issues in NSE experiments.

More specifically, vesicles suspensions of DOPC SUVs and G-C18:1 are prepared at pD 6 in $D_2O$ at concentration, $C_{DOPC}$= 10 mg/mL and $C_{G\text{-}C18:1}$= 10 mg/mL. 500 µL of each solution are mixed together for a total volume of 1 mL, final concentration of 5 mg/mL of each compound and molar ratio [DOPC]/[G-C18:1]= 0.6. modification of acidity is performed on this same sample to pD 3.6 first and to pD 10 afterwards. Despite the fact that the sample is prepared fresh, the intrinsically long acquisition times of NSE (3 h) do not allow a specific discussion on the possible morphological evolution in time. It is assumed here that the sample after 3 h is representative of a freshly prepared sample.

**Neutron spin echo (NSE) experiments: data analysis methodology**

To analyze the data, the Zilman-Granek (ZG) model[10] was applied. Starting from a Helfrich bending Hamiltonian,[11] the ZG model predicts a stretched exponential shape of the intermediate scattering function $S(q,t)$ (Eq. 1) with a stretch exponent $\beta$= 2/3, $I(q,t)$ being the spin-echo intensity at a given value of $q$ and time, $t$

$$\frac{I(q,t)}{I(q,0)} = S(q,t) = e^{-(\Gamma_{ZG}t)^{\beta}} \qquad \text{Eq. 1}$$

where

$$\Gamma_{ZG} = \alpha\gamma\left(\frac{k_bT}{k}\right)^{\frac{1}{2}}\left(\frac{k_bT}{\eta}\right)q^3 \qquad \text{Eq. 2}$$

from which, the scaled bending rigidity, $k$, is

$$\frac{k}{k_bT} = \left(\alpha\gamma \frac{k_bT}{\eta} \frac{q^3}{\Gamma_{ZG}}\right)^2 \qquad \text{Eq. 3}$$

where, $\alpha$= 0.0069 is a prefactor (commented below), $\gamma \approx 1$ for $\frac{k_bT}{k} << 1$, $\eta$ is the solvent viscosity (here taken as the viscosity of deuterated water, $\eta$= 0.00109 Pa.s at 25°C)[12], $k_b$ is the Boltzmann constant, $T$ is the temperature in Kelvin. The relaxation mode observed in the $q$ and $t$ range of NSE is not a pure bending mode[13] but a combined bending-stretching mode, which also depends on the compressibility modulus $k_c$. $k_c$ is proportional to the bending rigidity[14] and results in a renormalized bending rigidity, which can simply be used in the framework of the Zilman-Granek model,[15] thus resulting in a modification of the prefactor, $\alpha$, in Eq. 2. The exact value of $\alpha$ is still a matter of debate, but the current consensus seems to point towards a value of $\alpha$= 0.0069, which is different from 0.025 as originally suggested and it accounts for the fact that at the time and length scale of NSE a combined bending and stretching mode is observed.[13–17] The current consensus around the value of 0.0069 is also the result of matching the bending rigidity obtained by NSE (unrelaxed membrane rigidity), and pipette aspiration (bare bending rigidity) measurements.[18,19] In a recent review, Gupta et al.[20] give a comprehensive overview of the prefactors that were used by different groups.

In practice, this work plots the function $\frac{\Gamma_{ZG}}{q^3}$ as a function of $q$ and it employs a linear fit above ~0.06 Å$^{-1}$ with imposed zero slope, where it is virtually q-independent (Figure S 10b). At q< ~0.06 Å$^{-1}$, $\frac{\Gamma_{ZG}}{q^3}$ is not q-independent anymore and for this reason, the experimental points are not considered. Such deviations at low q are commonly observed[21] and they mark the limit of validity of the simple Zilman-Granek model, as other relaxation modes come into play as well.[21,22] Eventually, $\frac{\Gamma_{ZG}}{q^3}$ at each $q$-value are used in Eq. 3 to calculate the value of $\frac{k}{k_bT}$ at each $q$-value. By fitting the $\frac{k}{k_bT}(q)$ plot using with a linear function with slope imposed to zero (Figure S 10b), we could extract the value of $\frac{k}{k_bT}$ for each sample and reported in the manuscript (Figure 5).

**List of Supplementary Video files**

**Video S 0 and Video S 1** – Two experimental replicates showing the time-resolved *in situ* evolution of DOPC GUVs suspension to which an aliquot of G-C18:1 (5 mg/mL) vesicles suspension is added (experimental conditions: Table 1, experimental conditions in the article). Volume of sample, V= 96 µL, $V_{DOPC\ GUVs}$= 1 µL, $V_{G\text{-}C18:1}$= 1 µL, pH= 5.5, T= 23 ±1 °C, frequency of acquisition, f= 20 Hz. Red dye: Liss-PE; blue dye: perylene. Extracts of Video S 0 and Video S 1 are given in Figure 2 in the main text.

**Video S 2 and Video S 3** - Two experimental replicates showing the time-resolved *in situ* evolution of DOPC GUVs suspension to which an aliquot of G-C18:1 (0.5 mg/mL) vesicles suspension is added (experimental conditions: same as Figure 2 in Table 1, experimental conditions in the main article). Volume of sample, V= 96 µL, $V_{DOPC\ GUVs}$= 1 µL, $V_{G\text{-}C18:1}$= 1 µL, pH= 5.5, T= 23 ±1 °C, frequency of acquisition, f= 20 Hz. Red dye: Liss-PE; blue dye: perylene. Extracts of Video S 2 are given in Figure 2 in the main text.

**Video S 4** - Time-resolved *in situ* evolution of DOPC/G-C18:1 GUVs suspension (equilibration time: 1 h) to which a 0.1 M NaOH solution is added (experimental conditions: same as Figure 6 in Table 1, experimental conditions in the main article). Volume of sample, V= 96 µL, $V_{DOPC\ GUVs}$= 1 µL, $V_{G\text{-}C18:1}$= 1 µL, $C_{G\text{-}C18:1}$= 5 mg/mL, initial pH= 5.5, $V_{NaOH}$ = 1 uL, [NaOH]= 0.1 M, T= 23 ±1 °C, frequency of acquisition, f= 20 Hz. Red dye: Liss-PE; blue dye: perylene. Extracts of Video S 4 are given in Figure 6 in the main text.

**Video S 5** - Time-resolved *in situ* evolution of DOPC/G-C18:1 GUVs suspension (equilibration time: 1 h) to which a 1 M HCl solution is added (experimental conditions: same as Figure 7 in Table 1, experimental conditions in the main article). Volume of sample, V= 96 µL, $V_{DOPC\ GUVs}$= 1 µL, $V_{G\text{-}C18:1}$= 1 µL, $C_{G\text{-}C18:1}$= 5 mg/mL, initial pH= 5.5, $V_{HCl}$ = 1 uL, [HCl]= 1 M, T= 23 ±1 °C, frequency of acquisition, f= 20 Hz. Red dye: Liss-PE; blue dye: perylene. Extracts of Video S 5 are given in Figure 7a,b in the main text.

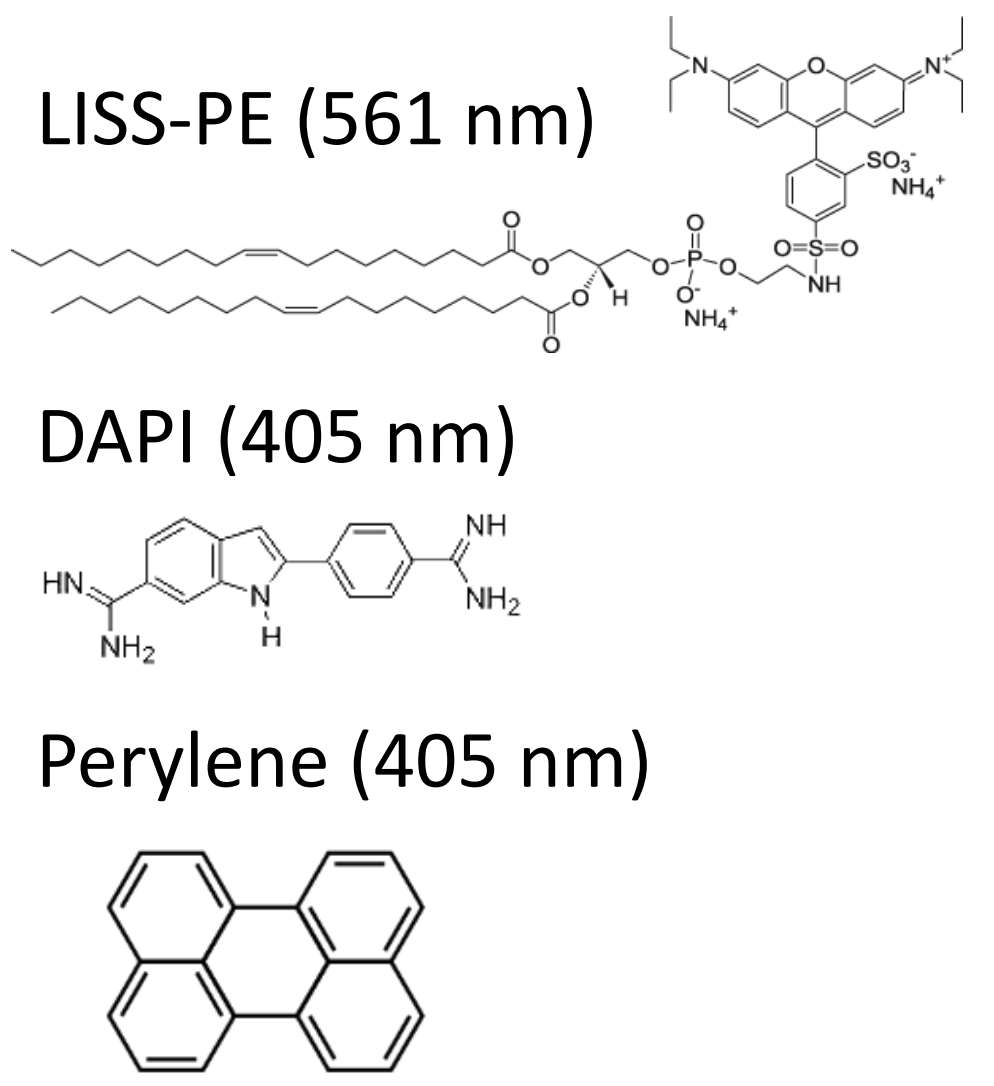

**Figure S 1 – Chemical structures of the chromatophores and $\lambda_{ex}$ used in this work**

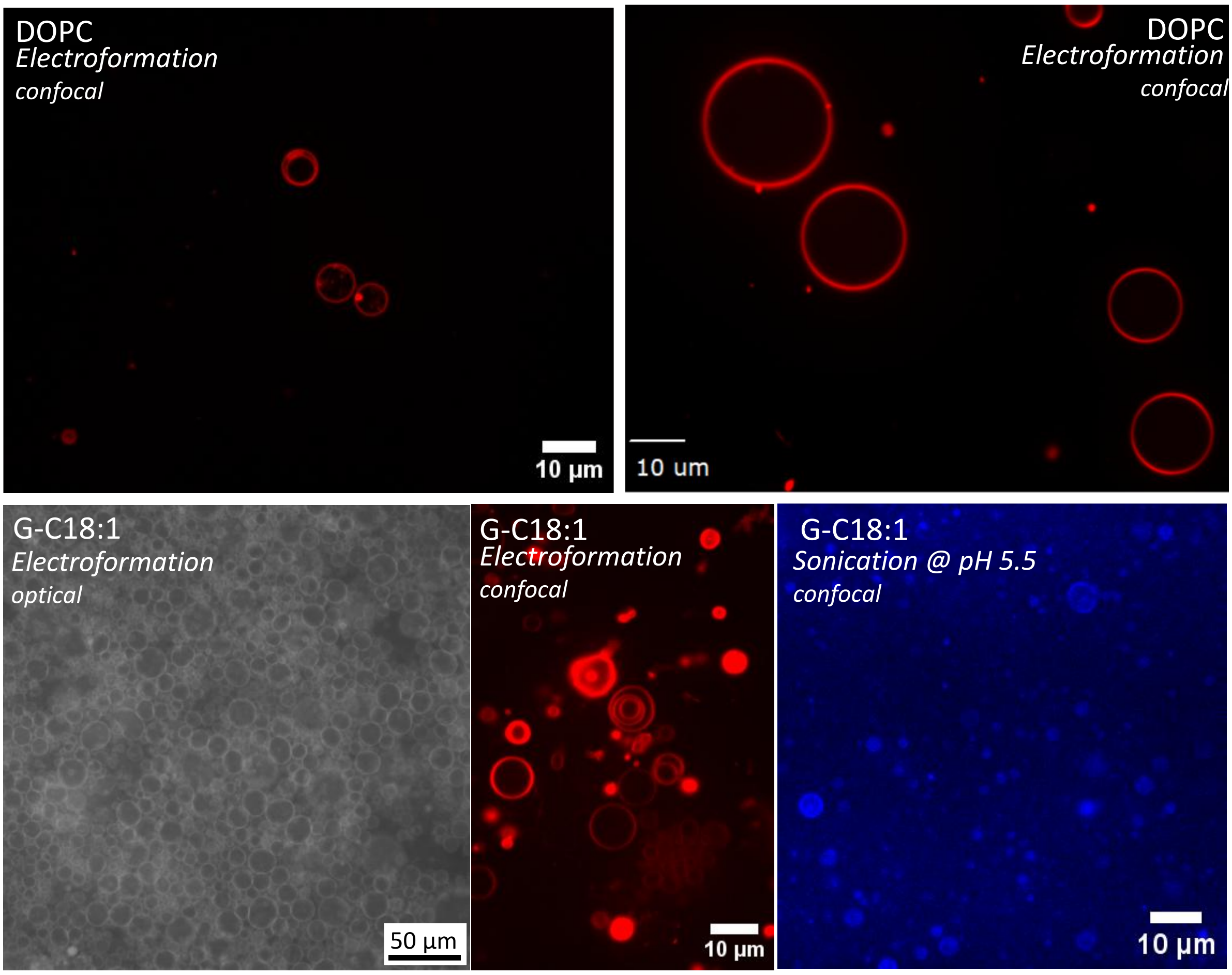

**Figure S 2 – Confocal and optical microscopy images recorded for control vesicles systems prepared from DOPC and G-C18:1. All images showing GUVs prepared by electroformation are stained with Liss-PE (red) while images from G-C18:1 vesicles prepared by sonication are stained with perylene (blue). The confocal micrograph recorded for G-C18:1 GUVs prepared by electroformation could only be obtained inside the electroformation cell itself by gently replacing the thick ITO slide with a standard thin glass slide.**

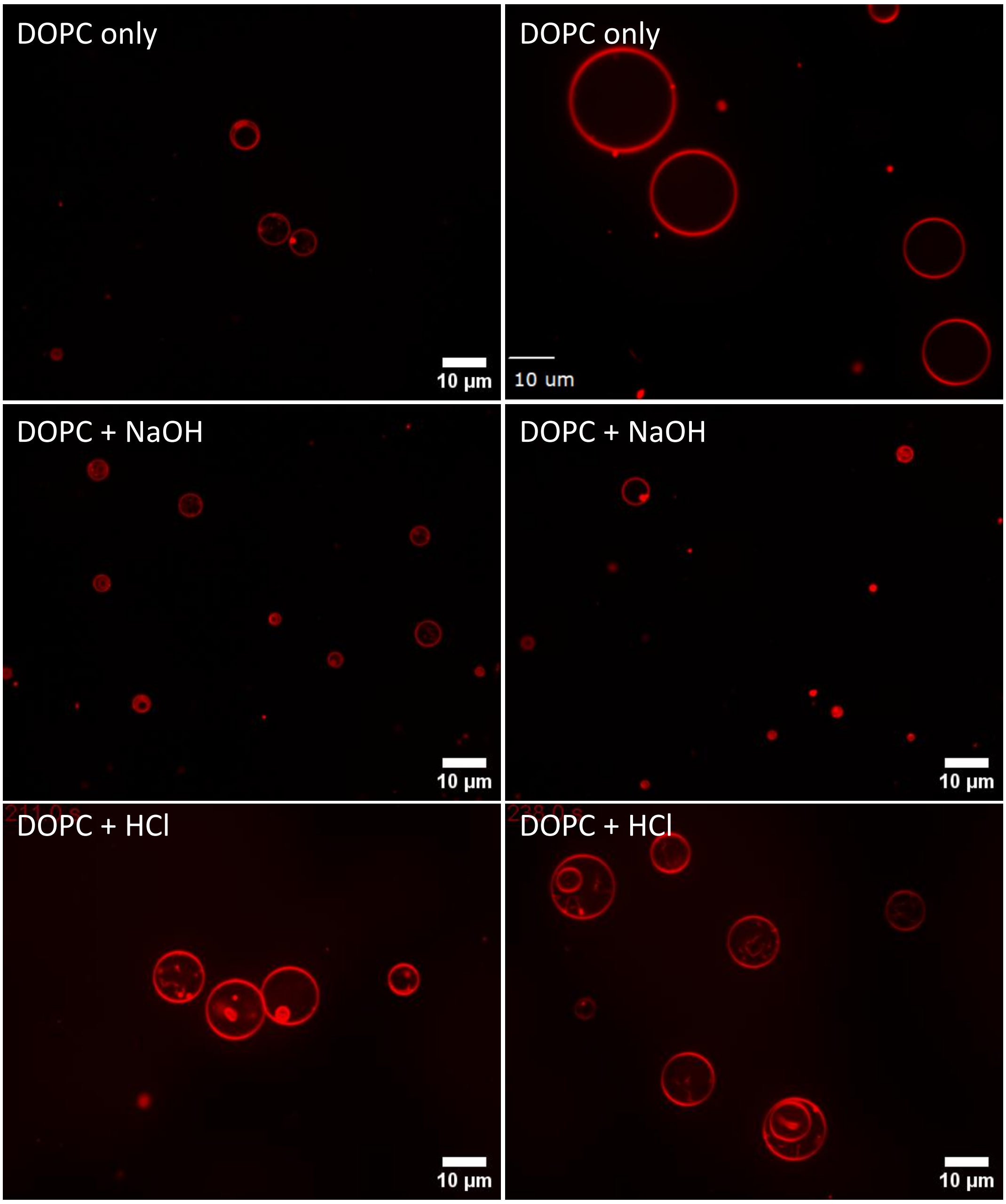

**Figure S 3 – Confocal images (dye: Liss-PE) of control GUVs prepared from DOPC by electroformation to which either NaOH (0.1 M) or HCl (1 M) is added. Conditions: $V_{DOPC}$= 1 µL; $V_{H2O}$= 96 µL; $V_{NaOH}$= 2 µL; $V_{HCl}$= 1 µL. To promote sedimentation of GUVs, DOPC contains 0.1 M of a sucrose solution, while the aqueous medium contains a 0.1 M glucose solution.**

**Additional data on the interactions between DOPC GUVs and G-C18:1 vesicles**

Figure S 4 to Figure S 8 show a series of complementary experiments, which aim at demonstrating the interaction between the DOPC GUVs and G-C18:1 vesicles. This is done by a set of three different experiments: 1) colocalization of red (from DOPC) and blue (from G-C18:1) dyes (Figure S 4, Figure S 5), 2) the interaction between a positively-charged blue dye (DAPI) with the negatively-charged G-C18:1 (Figure S 6, Figure S 7) and 3) osmotic stress (Figure S 8).

Compared to a G-C18:1-free control, Figure S 4a,b show that rhodamine (carried by DOPC GUVs) and perylene (carried by G-C8:1 vesicles) are co-localized. However, the distribution of perylene seems to be heterogeneous across GUVs, with variations ranging from few % to 50% of the perylene reference signal (Figure S 5). Further proof is generated by two additional experiments. In the first one, it is supposed that DAPI (Figure S 1), a positively charged (at pH 5.5), water-soluble, blue dye, interacts by electrostatic attraction with DOPC GUVs rich in G-C18:1, negatively charged at pH 5.5 (pKa= 5.69).[23] For this specific experiment, to avoid color conflicts, G-C18:1 vesicles are not labeled with perylene. Both the action of adding DAPI to a DOPC/G-C18:1 mixed GUV system (Figure S 6a) and adding G-C18:1 to DAPI-labeled DOPC GUVs (Figure S 6b-d, Figure S 7) show an increase in the blue color of the mixed DOPC/G-C18:1 system compared to the staining of DOPC in the absence of G-C18:1.

A third strategy to demonstrate fusion employed the difference in dye (LISS-PE) partitioning between liquid ordered ($L_o$) and liquid disordered ($L_d$) phases in binary and ternary mixtures, whereas LISS-PE prefers the $L_d$ phase:[24,25] the control DOPC:DPPC:cholesterol (2:2:1) indeed shows phase separated domains (Figure S 8a), whereas DPPC (Tm ~ 41 °C) is in the solid-like gel ($L_\beta$) state at room temperature. Phase separation can also be observed in the ternary DOPC/cholesterol/G-C18:1 system, where G-C18:1 is added to DOPC/cholesterol (1:1) GUVs (Figure S 8b). Applying an osmotic stress to the ternary system enhances the membrane phase separation process between $L_o$ and $L_d$ phases, as expected for model membranes,[26] whereas the colocalization of red and blue dyes further confirm the fusion (Figure S 8c). Observing phase separation in the DOPC/cholesterol/G-C18:1 system further confirms the insertion of G-C18:1 in the DOPC membrane.

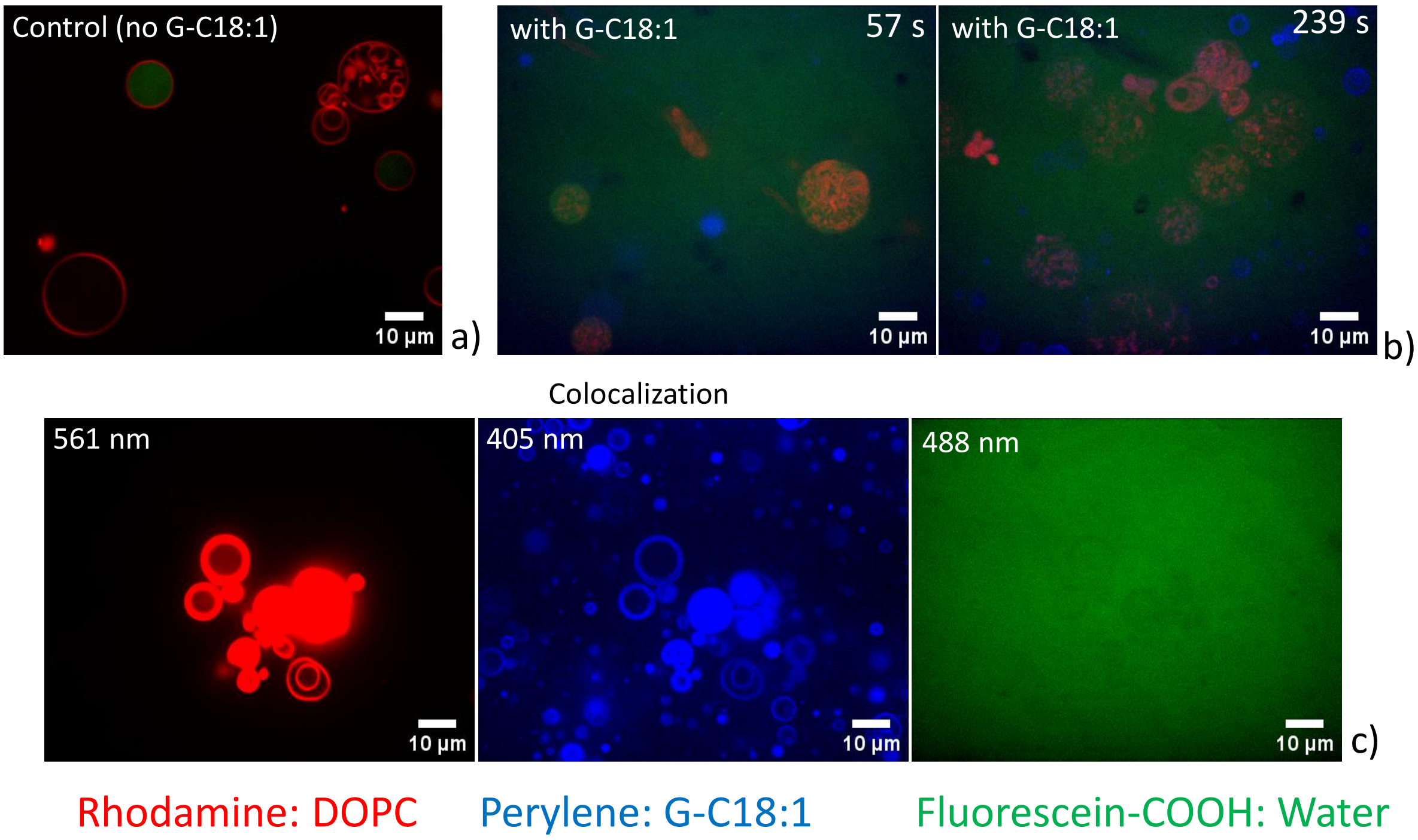

**Figure S 4 – Confocal images (dyes: Red: Liss-PE; Blue: perylene; Green: fluorescein) of a) control DOPC (2 mg/mL) GUVs prepared by electroformation ($V_{DOPC}$= 1 µL; V= 96 µL, pH 5.5) and b) G-C18:1 (5 mg/mL) vesicles added to DOPC GUVs ($V_{G\text{-}C18:1}$= 1 µL). Time scale refers to time elapsed after addition. c) Colocalization: confocal image split into red, blue and green channels.**

The experiment in Figure S 4 demonstrates colocalization of red (DOPC) and blue (G-C18:1) dyes. Fluorescein (green), initially contained in DOPC vesicles, leaks out of the GUVs after contact between DOPC and G-C18:1. The green color eventually spreads in the entire volume of the solution, thus demonstrating membrane damage.

More explanatory text is given on Page S 12.

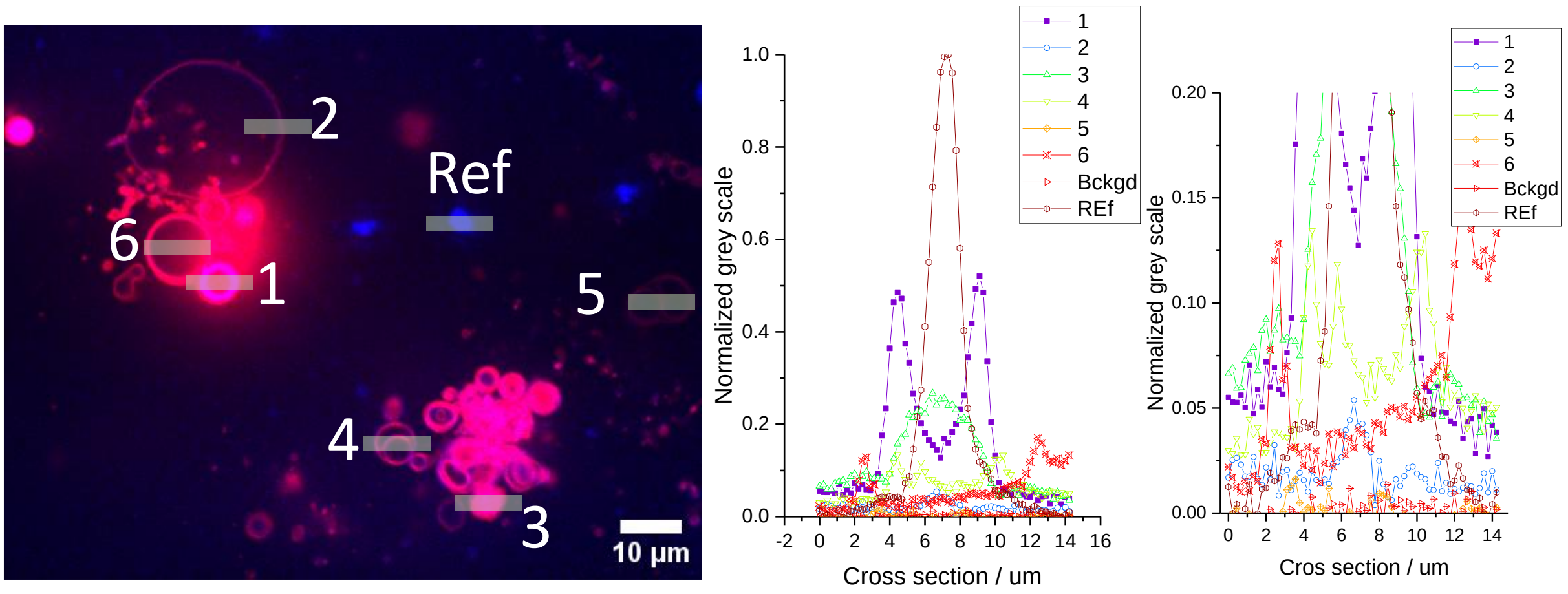

**Figure S 5 – Confocal image (dyes: Red: Liss-PE) of DOPC GUVs (electroformation, $V_{\text{DOPC/sucrose 0.1 M}}$= 1 µL; $V_{\text{glucose 0.1 M}}$= 96 µL) containing unlabeled (no perylene) G-C18:1 vesicles ($V_{\text{G-C18:1 5 mg/mL / sucrose 0.1 M}}$= 1 µL). DAPI (405 nm) at pH 5.5 is added to the mixture. The image is recorded about 1 h after mixing. Right-hand side: intensity profiles measured on the confocal image (grey bars) and corresponding to the DAPI channel only. Normalization is performed on the intensity of the blue region, here taken as reference, and containing G-C18:1 vesicles only.**

DAPI is employed instead of perylene as this dye, at acidic pH, is positively charged and it is supposed to strongly interact with G-C18:1, partially negatively charged. In Figure S 5, DAPI is added to the mixed DOPC/G-C18:1 system and increase in intensity at 405 nm (blue) is evaluated. The strong evolution in the cross-sectional intensities shows that the dye distribution is not uniform from one vesicle to another.

More explanatory text is given on Page S 12.

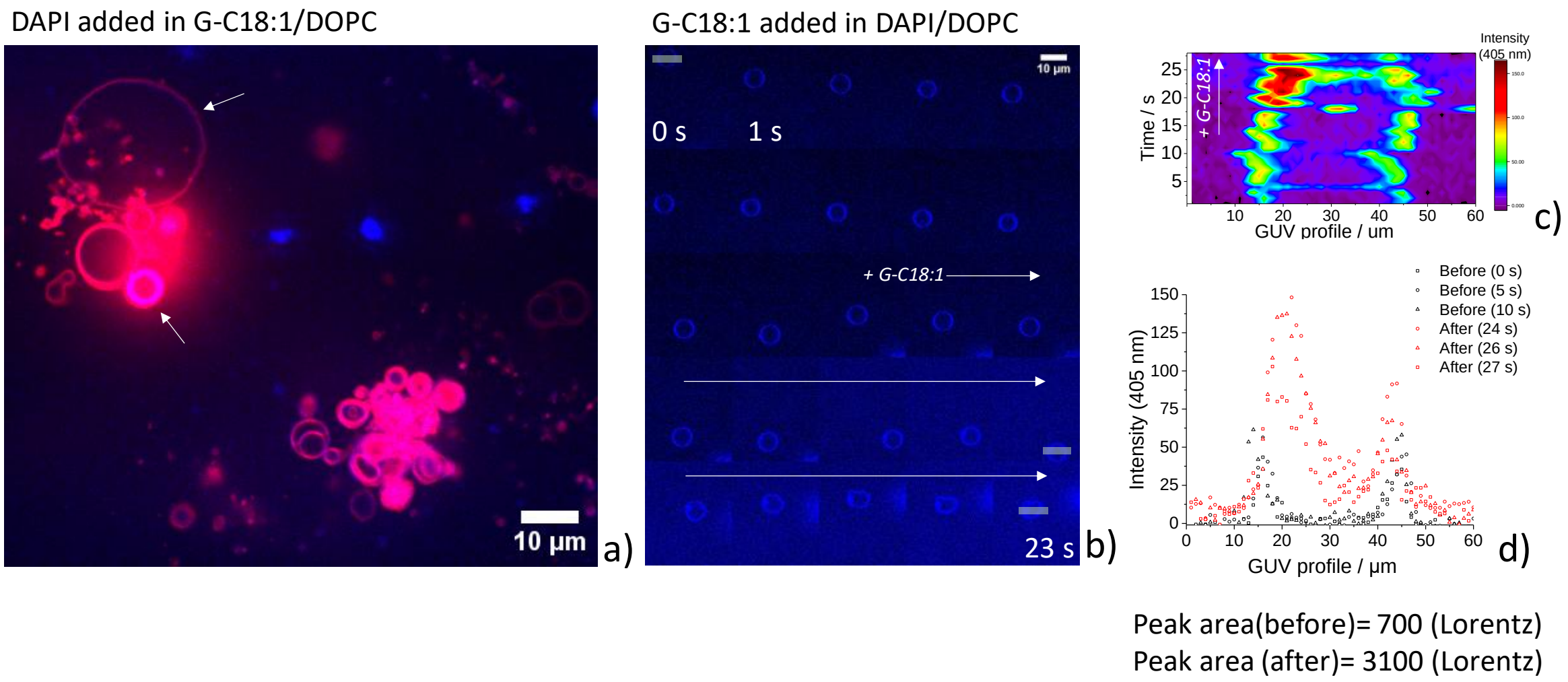

**Figure S 6 – a) Confocal image (dyes: Red: Liss-PE; blue: DAPI) of DOPC GUVs (electroformation, $V_{DOPC/sucrose\ 0.1\ M}$= 1 µL; $V_{glucose\ 0.1\ M}$= 96 µL) containing unlabellaed (no perylene) G-C18:1 vesicles ($V_{G\text{-}C18:1\ 5\ mg/mL\ /\ sucrose\ 0.1\ M}$= 1 µL). DAPI (405 nm) at pH 5.5 is added to the mixture. The image is recorded about 1 h after mixing. b) G-C18:1 ($V_{G\text{-}C18:1\ 5\ mg/mL\ /\ sucrose\ 0.1\ M}$= 1 µL) is added to a DOPC GUVs solution (electroformation, $V_{DOPC/sucrose\ 0.1\ M}$= 1 µL; $V_{glucose\ 0.1\ M}$= 96 µL) already containing DAPI. The intensity of DAPI (405 nm) is followed with time. The intensity profile of the GUV is collected along the grey bar at all recorded times and reported in c). c-d) Intensity profiles of the GUV shown in b) recorded with time. Time-dependent contour is given in (c) while selected profiles before and after addition of G-C18:1 are given in d) to show the increase in the intensity.**

Despite the stronger interaction between DAPI and G-C18:1, a background interaction occurs between DAPI and DOPC GUVs. The experiment in Figure S 6 is performed with the goal of separating the contribution of DAPI adsorption to the DOPC GUVs from the contribution of DAPI adsorption due to G-C18:1. DAPI is added to a G-C18:1/DOPC system (Figure S 6a) and to a DOPC GUV system alone, to which G-C18:1 is eventually added (Figure S 6b). The increase in the DAPI (blue, 405 nm) intensity with time after addition of G-C18:1 is shown in Figure S 6c,d.

More explanatory text is given on Page S 12.

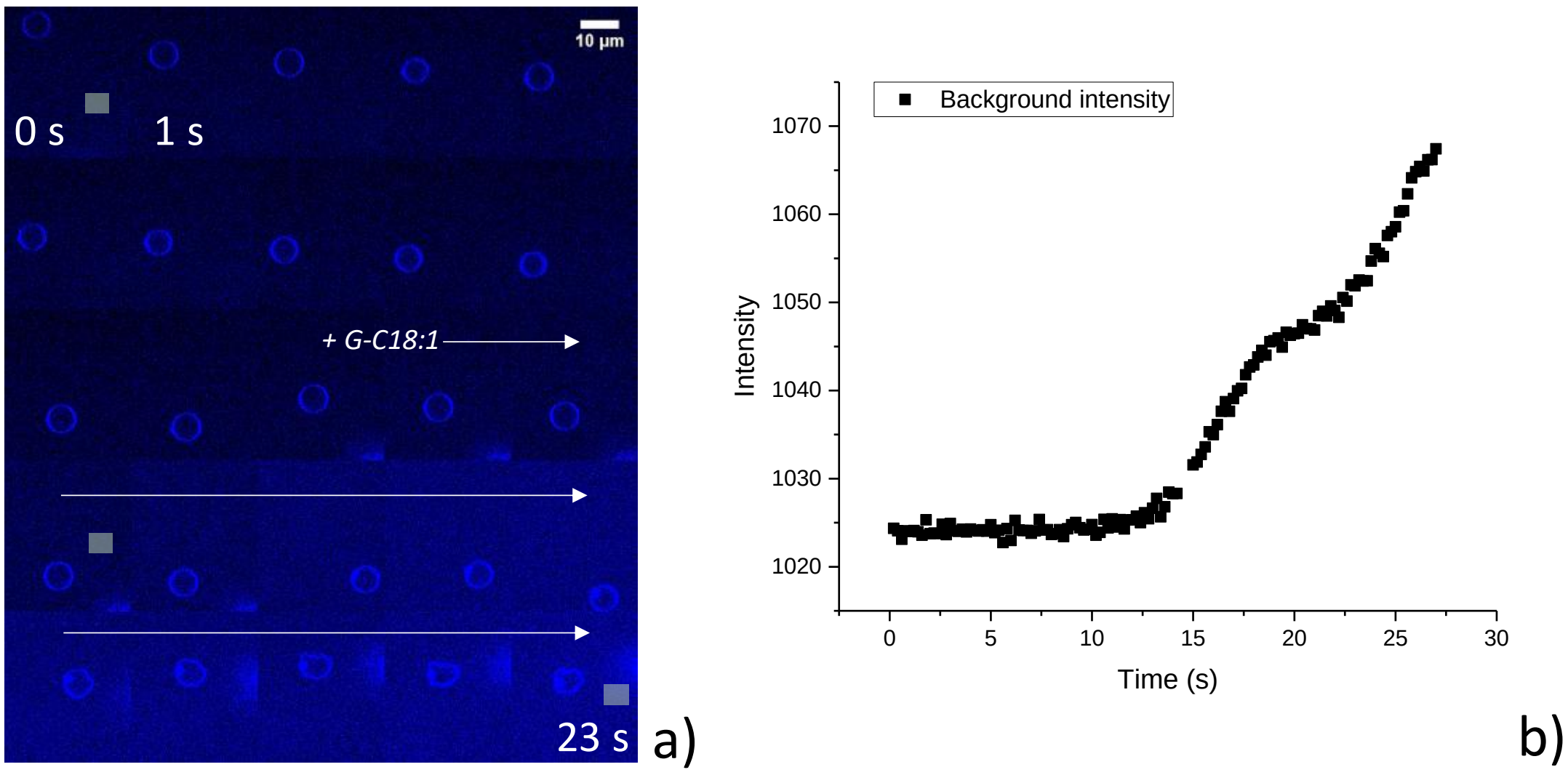

**Figure S 7 – a,b) Evolution of the background signal (blue channel, 405 nm) of the experiment shown in Figure S 6b. The grey bar in the GUV panel in a) (lowest right-hand image) shows the location where the background signal is averaged at a given point in time and plot in b).**

Explanatory text is given on Page S 12.

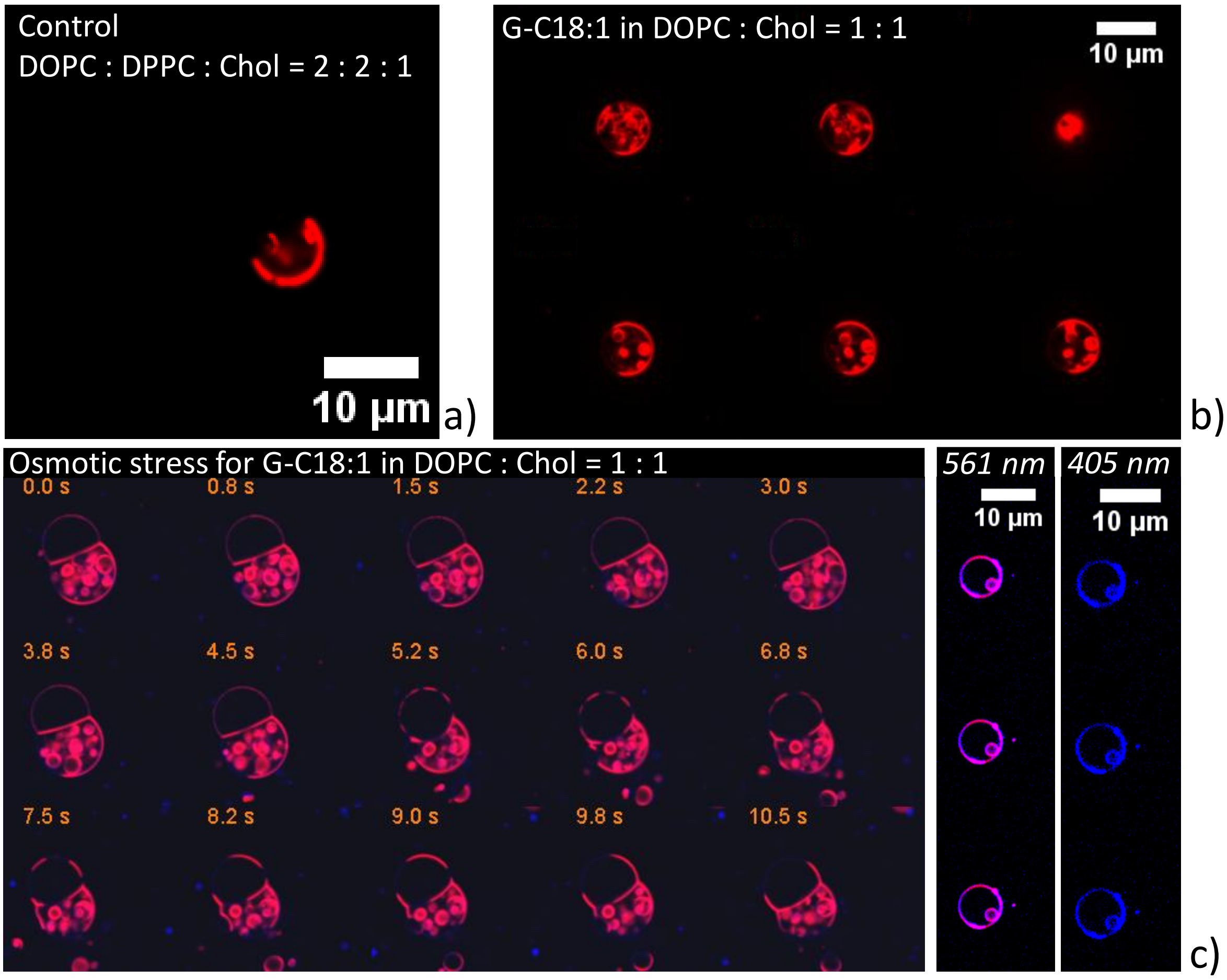

**Figure S 8 – a) Control experiment showing a confocal image (dye: Red: Liss-PE) of DOPC GUVs prepared from a 2:2:1 mixture of DOPC, DPPC and cholesterol (electroformation, $V_{DOPC/DPPC/Chol/sucrose\ 0.1\ M}$= 2 µL; molar ratio DOPC/DPPC/Chol= 2:2:1, $V_{glucose\ 0.1\ M}$= 96 µL). b) Confocal image of perylene-labelled G-C18:1 vesicles ($V_{G\text{-}C18:1\ 5\ mg/mL\ /\ sucrose\ 0.1\ M}$= 1 µL) added to a 1:1 suspension of DOPC and cholesterol GUVs ($V_{DOPC/Chol/sucrose\ 0.1\ M}$= 1 µL; molar ratio DOPC/Chol= 2:1, $V_{glucose\ 0.1\ M}$= 96 µL). c) Osmotic stress (unbalancing the inner and outer concentration of sucrose/glucose) experiment performed on b). The solution in b) is diluted by a factor 2 (added volume of glucose-free $H_2O$ is 100 µL), thus lowering the concentration of glucose to 0.05 M. In c), colocalization of the dyes (Liss-PE, perylene) is also shown for at selected location.**

Explanatory text is given on Page S 12.

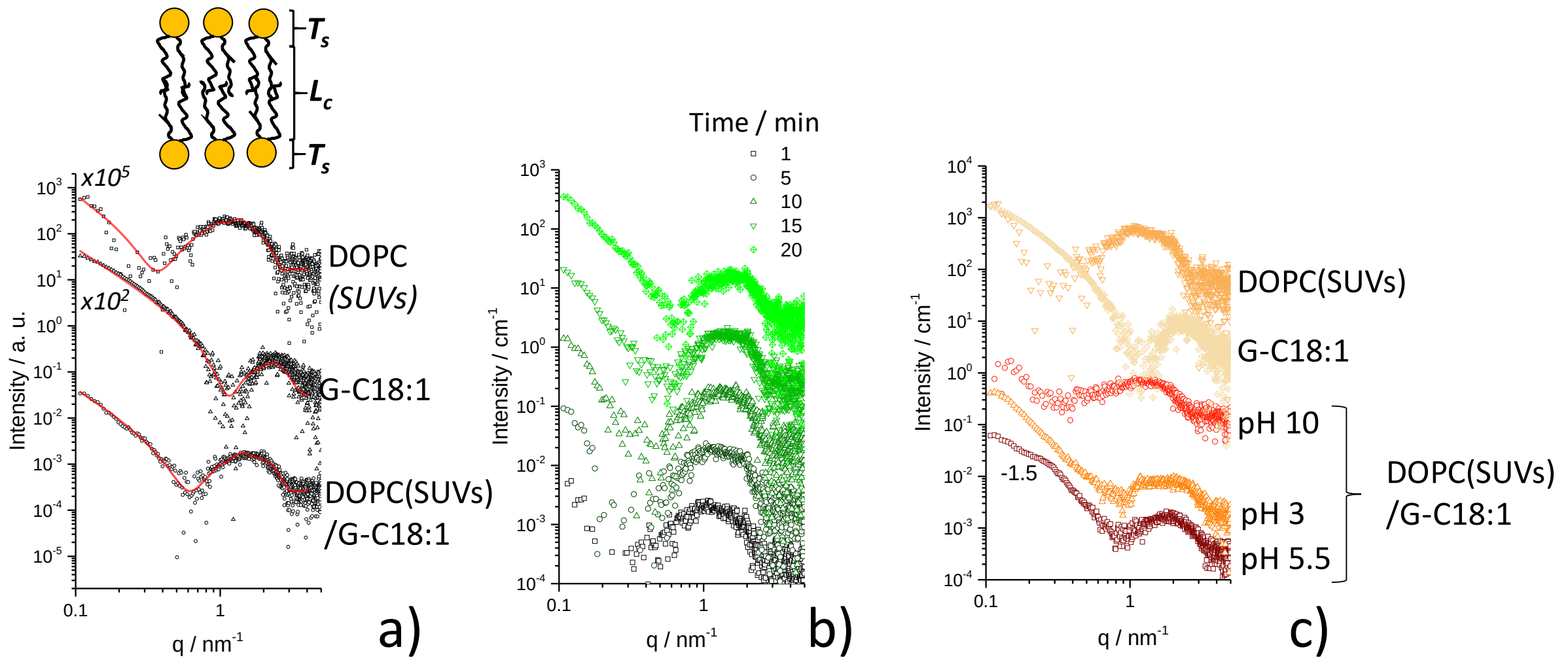

**Figure S 9 – Experiment performed on single unilamellar vesicles (SUVs). a) Static SAXS experiments recorded on G-C18:1 vesicle suspension (5 mg/mL, pH 5.5), DOPC SUVs (2 mg/mL, pH 5.5) and DOPC(SUVs)/G-C18:1 mixed molar ratio [DOPC]:[G-C18:1]= 1.2 ($C_{DOPC}$= 2 mg/mL, $C_{G\text{-}C18:1}$= 0.95 mg/mL, pH 5.5, refer to Table S 1 for more details). b) I(q)-representation of SAXS spectra recorded during time-resolved *in situ* SAXS experiment: aliquots of the G-C18:1 vesicle suspension are added every 5 min to the DOPC SUVs suspension (2 mg/mL, pH 5.5). The molar ratio [DOPC]:[G-C18:1] evolves from 5.9 at t= 0 to 1.2 at t= 20 min. All details on the sample preparation are given in Table S 1. c) Static SAXS experiments performed on DOPC(SUVs)/G-C18:1 at [DOPC]:[G-C18:1]= 0.6 at pH 5.5, pH 3 and pH 8. $C_{G\text{-}C18:1}$= 5 mg/mL, $C_{DOPC}$= 2 mg/mL. Details on the sample preparation are given in Table S 1.**

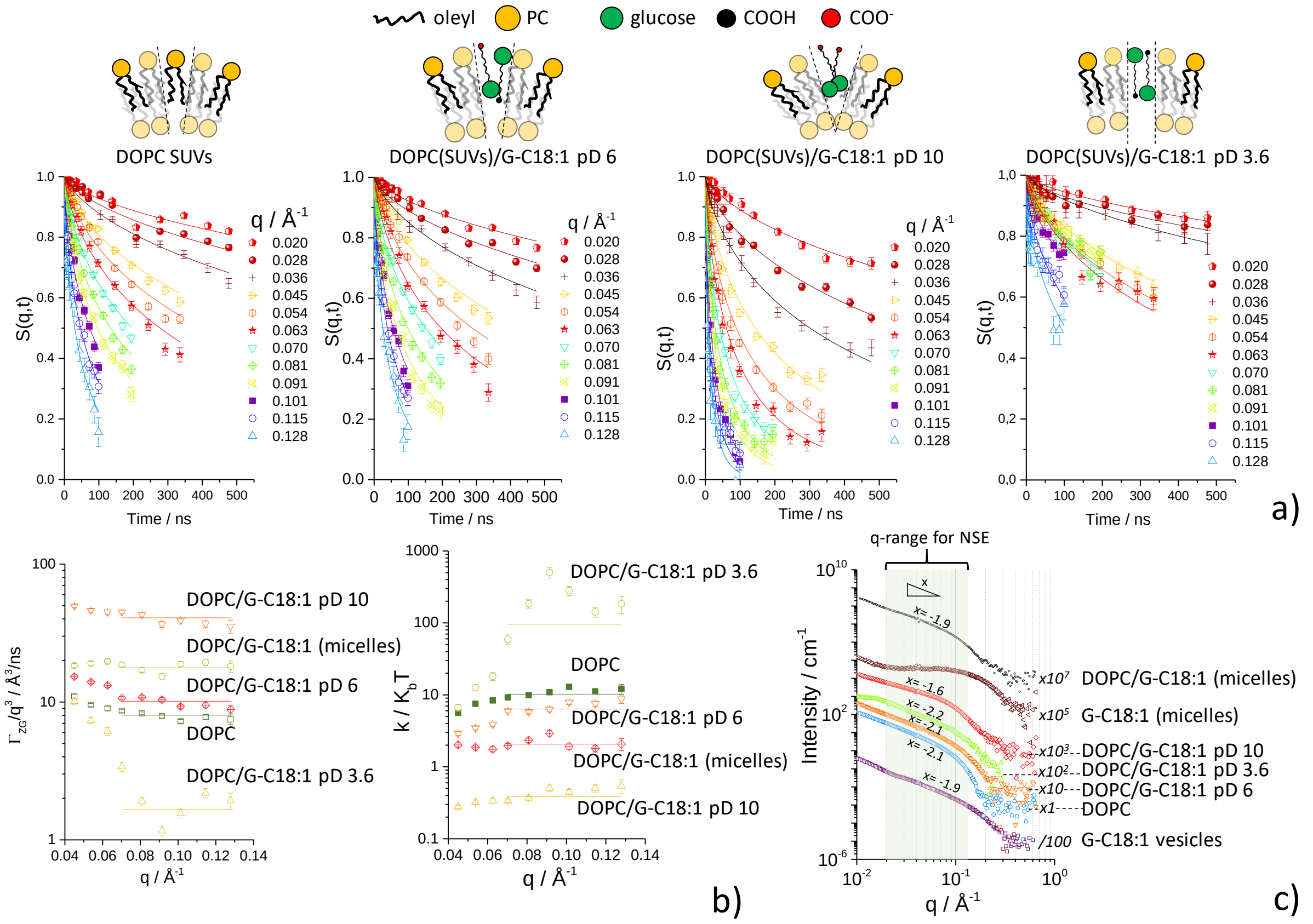

**Figure S 10 – a) Normalized, q-dependent, spin-echo intermediate scattering functions for DOPC, DOPC/G-C18:1 at pD 6, DOPC/G-C18:1 at pD 10 and DOPC/G-C18:1 at pD 3.6. The molar ratio [DOPC]/[G-C18:1]= 0.6. The samples at pD 3.6 and pD 10 were prepared by increasing or decreasing the pD of the sample at pD 6. b) Plots of the Zilman-Granek decay parameter, $\frac{\Gamma_{ZG}}{q^3}(q)$ (Eq. 2) and scaled bending rigidity, $\frac{k}{k_bT}(q)$ (Eq. 3) calculated from the intermediate scattering functions. The high-q portion of the plots is fitted with a linear function of imposed zero slope. The values of $\frac{k}{k_bT}(q=0)$ are reported in Figure 5 in the main text. c) Scaled SANS experiments (scaling factor next to plots) corresponding to the samples in a) but also including selected controls: G-C18:1 vesicles (pD 6, 10 mg/mL), G-C18:1 micelles (pD 10, 10 mg/mL), DOPC/G-C18:1 (micelles) (pD 10) (G-C18:1 micelles added to DOPC SUVs, [DOPC]/[G-C18:1]= 0.6, $C_{DOPC}$= $C_{G\text{-}C18:1}$= 5 mg/mL). All DOPC samples are SUVs and all experiments are performed in deuterated water (detailed information in the experimental section on neutron scattering).**

The intermediate scattering functions, S(q,t) are presented in Figure S 10a for a series of samples: DOPC SUVs, DOPC/G-C18:1 pD 6 and derived samples with modified pD (3.6 and 10). The fits of the intermediate scattering function using Eq. 1 provide the wavevector independent $\frac{\Gamma_{ZG}}{q^3}$ quantity (Eq. 2), from which the scaled bending rigidity is obtained from Eq. 3. $\frac{\Gamma_{ZG}}{q^3}$ and $\frac{k}{k_bT}$ are plotted against $q$ in Figure S 10b. The lower values of $\frac{\Gamma_{ZG}}{q^3}$ corresponding to the sample at pD 3.6 are coherent with the slower decay of its corresponding S(q,t) and indicative of an increase in the membrane stiffness compared to the parent sample at pD 5.5. Figure S 10b also shows the method (linear fit with imposed zero slope) to estimate $\frac{\Gamma_{ZG}}{q^3}$ and $\frac{k}{k_b T}(q = 0)$ (Eq. 3, with pre-factor 0.0069) values given in the main text (Figure 5).

Finally, Figure S 10c shows the typical SANS profiles recorded on the samples analyzed by NSE and selected controls. The SANS data are analyzed using a model-independent approach at q< 0.1 Å$^{-1}$. The I(q) power law in log-log scale expected for flat membranes, as found at this length scale for vesicles, is -2. The typical power laws, noted as $x$ next to each SANS profile in Figure S 10c, are contained between -1.9 and -2.1 for most samples, thus indicating that G-C18:1 vesicles (pD 6), DOPC SUVs (pD 6), DOPC/G-C18:1 mixed vesicles (pD 6) and DOPC/G-C18:1 (pD 3.6) are all characterized by a 2D membrane structure.[27] Interestingly, the control sample, DOPC SUVs to which a G-C18:1 micellar solution (pD 10) is added also displays a power law of -1.9, indicating a 2D membrane structure for this sample, as well. The SANS profile of a micellar solution of G-C18:1 at pD 10, also shown in Figure S 10c, is on the contrary typical of spheroidal micelles undergoing repulsive interactions.[28,29] The 2D membrane structure suggested by SANS indicates that the evolution in $\frac{k}{k_b T}$ measured on the samples in Figure S 10 can be reasonably explained by an actual change in their corresponding membrane rigidity.

The only exception seems to concern the sample associated to DOPC/G-C18:1 at pD 10. The low-q slope seems to be characterized by a different power law, of about -1.6. This value can be explained either by the presence of wormlike micelles,[30,31] or by the coexistence of objects of different morphology (spheroidal micelles and vesicles). Caution should then be taken in regard to the corresponding scaled bending rigidity, $\frac{k}{k_b T}$.

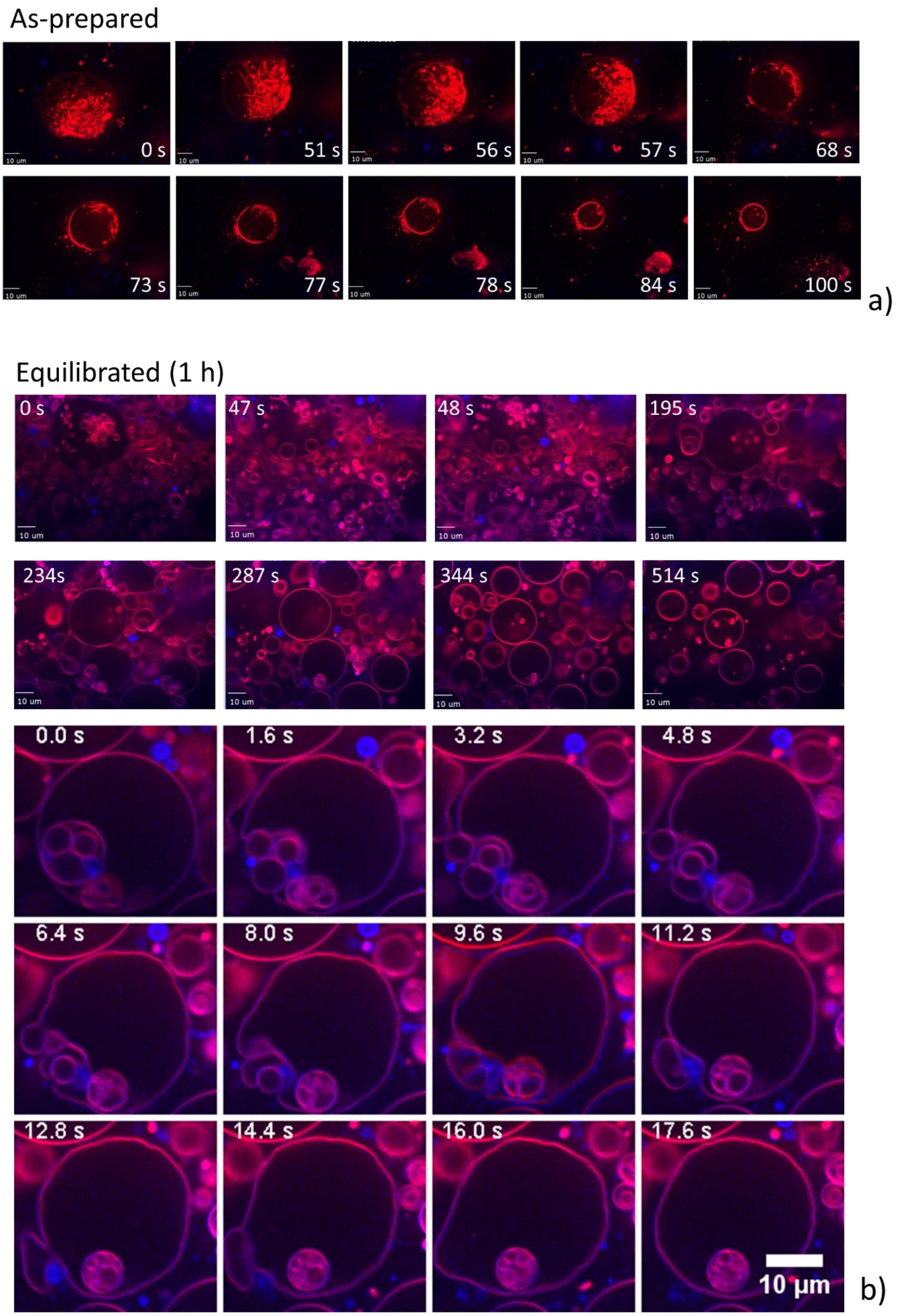

**Figure S 11 – Additional images of the time-resolved *in situ* evolution of a) as-prepared and b) equilibrated (1 h) DOPC/G-C18:1 GUVs suspension to which a 0.1 M NaOH (1 µL) solution is added. Experimental conditions prior of NaOH addition: V= 96 µL, $V_{DOPC\ GUVs}$= 1 µL, $V_{G\text{-}C18:1}$= 1 µL, $C_{G\text{-}C18:1}$= 5 mg/mL, T= 23 ±1 °C, pH 5.5, frequency of images acquisition, f= 20 Hz. To promote sedimentation of GUVs, DOPC contains 0.1 M of a sucrose solution, while the aqueous medium contains a 0.1 M glucose solution.**

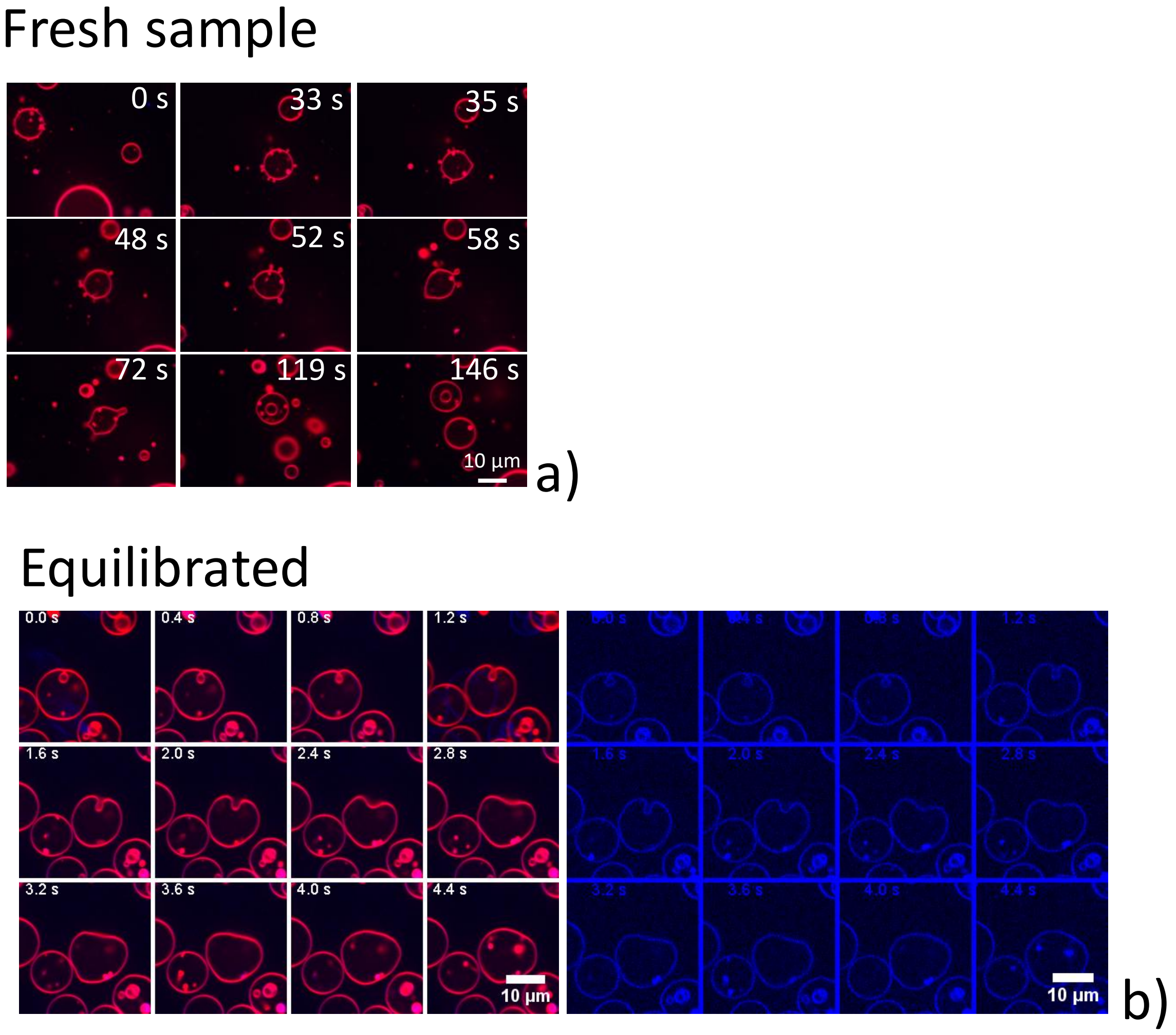

**Figure S 12 – Time-resolved *in situ* evolution of a) fresh and b) equilibrated (1 h) DOPC/G-C18:1 GUVs suspension to which a 0.01 M NaOH solution (3 µL) is added. The concentration of G-C18:1 and added base has been divided by a factor 10 compared to Figure S 11. Colocalization between the blue and red channels are shown in the equilibrated set of images. Experimental conditions prior of NaOH addition: V= 96 µL, $V_{DOPC\ GUVs}$= 1 µL, $V_{G\text{-}C18:1}$= 1 µL, $C_{G\text{-}C18:1}$= 0.5 mg/mL, T= 23 ±1 °C, pH 5.5, frequency of images acquisition, f= 20 Hz. To promote sedimentation of GUVs, DOPC contains 0.1 M of a sucrose solution, while the aqueous medium contains a 0.1 M glucose solution.**

# Effect of G-C18:1 micelles on DOPC GUVs

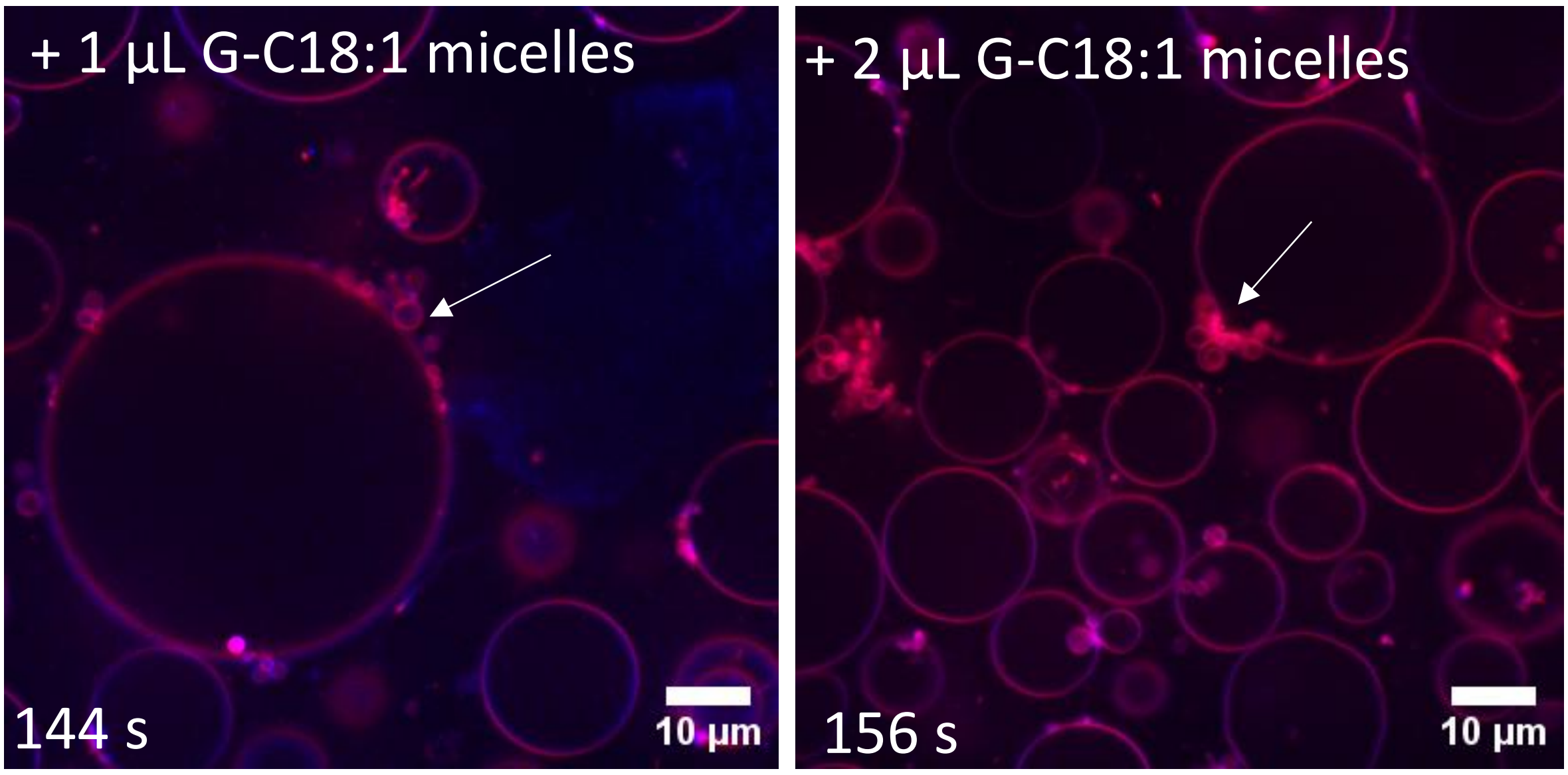

**Figure S 13 – Control. Time-resolved *in situ* evolution of a DOPC GUVs suspension (pH 8) to which a G-C18:1 micellar solution prepared at pH 8 is added. Experimental conditions: V= 96 µL, $V_{DOPC\ GUVs}$= 1 µL, $V_{G\text{-}C18:1}$= 2 µL, $C_{G\text{-}C18:1}$= 5 mg/mL, T= 23 ±1 °C, pH 8, frequency of images acquisition, f= 20 Hz. DOPC is labeled with LISS-PE and the G-C18:1 micellar solution with perylene. To promote sedimentation of GUVs, DOPC contains 0.1 M of a sucrose solution, while the aqueous medium contains a 0.1 M glucose solution.**

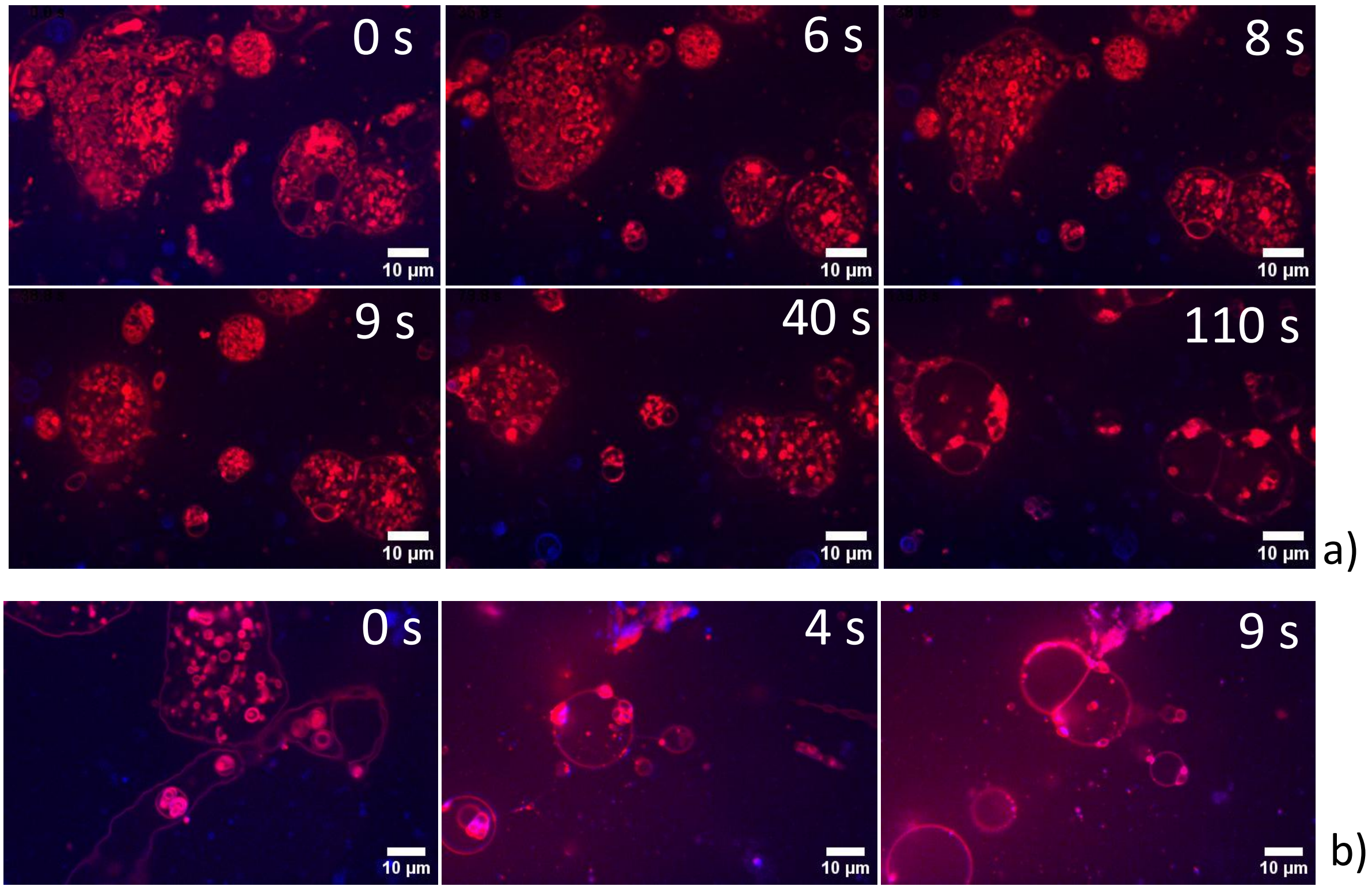

**Figure S 14 – a,b) Time-resolved *in situ* experiment showing the evolution of an equilibrated (1 h) DOPC/G-C18:1 GUVs suspension to which a 1 M HCl solution (1 µL) is added. The experiment in a) is repeated twice. The second experiment is shown in b). Experimental conditions prior of HCl addition: V= 96 µL, $V_{DOPC\ GUVs}$= 1 µL, $V_{G\text{-}C18:1}$= 1 µL, $C_{G\text{-}C18:1}$= 5 mg/mL, T= 23 ±1 °C, pH 5.5, frequency of images acquisition, f= 20 Hz.**